\documentclass[traditabstract]{aa}
\usepackage{txfonts}
\usepackage{graphicx}
\usepackage{natbib}
\renewcommand{\deg}{\mbox{$^{\circ}$}~}
\newcommand{\kms}{\mbox{km~s$^{-1}$}}
\newcommand{\mols}{\mbox{molec.~s$^{-1}$}}

\newcommand{\psj}{\textit{Planetary Science Journal}}

\begin{document}

\title{Molecular composition of comet 46P/Wirtanen from millimetre-wave
  spectroscopy
\thanks{Based on observations carried out with the IRAM-30m and NOEMA telescopes.
IRAM is supported by INSU/CNRS (France), MPG (Germany), and IGN (Spain).}
\thanks{The radio spectra are available at the CDS via anonymous 
ftp to cdsarc.u-strasbg.fr (130.79.128.5)
or via http://cdsweb.u-strasbg.fr/cgi-bin/qcat?J/A+A/.}
}
 
\author{N. Biver\inst{1}
   \and D. Bockel\'ee-Morvan\inst{1}
   \and J. Boissier\inst{2}
   \and R. Moreno\inst{1}
   \and J. Crovisier\inst{1}
   \and D.C. Lis\inst{3}
   \and P. Colom\inst{1}
   \and M. Cordiner\inst{4,5}
   \and S. Milam\inst{4}
   \and N.X. Roth\inst{4} 
   \and B.P. Bonev\inst{6}
   \and N. Dello Russo\inst{7}
   \and R. Vervack\inst{7}
   \and M. A. DiSanti\inst{8}
   }
\institute{LESIA, Observatoire de Paris, PSL Research University, CNRS, 
  Sorbonne Universit\'e, Universit\'e de Paris,
  5 place Jules Janssen, F-92195 Meudon, France
 \and IRAM, 300, rue de la Piscine, F-38406 Saint Martin d'H\`eres, France
 \and Jet Propulsion Laboratory, California Institute of Technology,
      4800 Oak Grove Drive, Pasadena, CA, 91109, USA 
 \and Solar System Exploration Division, Astrochemistry Laboratory Code 691,
      NASA-GSFC, Greenbelt, MD 20771, USA
 \and Department of Physics, Catholic University of America,
      Washington, DC 20064, USA
 \and Department of Physics, American University, Washington, DC, USA 
 \and Johns Hopkins University Applied Physics Laboratory,
      11100 Johns Hopkins Rd., Laurel, MD 20723, USA
 \and Solar System Exploration Division, Planetary System Laboratory Code 693,
      NASA-GSFC, Greenbelt, MD 20771, USA
}

   \titlerunning{Composition of comet 46P/Wirtanen}
   \authorrunning{Biver et al.}
   \date{\today}

   \abstract{We present the results of a molecular survey of comet
     46P/Wirtanen undertaken with the IRAM 30-m and NOEMA radio telescopes
     in December 2018. Observations at IRAM 30-m during the 12--18 Dec.
     period comprise a 2~mm spectral survey covering 25~GHz and a 1~mm
     survey covering 62~GHz.
     The gas outflow velocity and kinetic temperature have been accurately
     constrained by the observations. We derive
     abundances of 11 molecules, some being identified remotely for the first
     time in a Jupiter-family comet, including complex
     organic molecules such as formamide, ethylene glycol, acetaldehyde,
     or ethanol. Sensitive upper limits on the abundances of
     24 other molecules are obtained.
     The comet is found to be relatively rich in methanol (3.4\% relative
     to water), but relatively depleted in CO, CS, HNC, HNCO, and HCOOH.}

   \keywords{Comets: general
-- Comets: individual:  46P/Wirtanen
-- Radio lines: planetary system -- Submillimeter: planetary system}

\maketitle

\section{Introduction}
Comets are the most pristine remnants of the formation of the
Solar System 4.6 billion years ago. They sample some of the oldest and most
primitive material in the Solar System, including ices, and are
thus our best window to the volatile composition of the solar
proto-planetary disk. Comets may also have played a role in the
delivery of water and organic material to the early Earth
\citep[see][and references therein]{Har11}.
The latest simulations of the early Solar System's evolution \citep{Bra13,Obr14}
suggest a more complex scenario. On the one hand, ice-rich bodies formed beyond
Jupiter may have been implanted in the outer asteroid belt and participated in
the supply of water to the Earth, or, on the other hand, current comets coming
from either the Oort Cloud or the scattered disk of the Kuiper belt may have
formed in the same trans-Neptunian region, sampling the same diversity of 
formation conditions. Understanding the diversity in composition and isotopic
ratios of comet material is thus essential in order to assess such
scenarios \citep{Alt03,Boc15}.  

Comet 46P/Wirtanen is a Jupiter-family comet (JFC) orbiting the Sun in 5.4 years
on a low inclination (11.7\deg) orbit. It reached perihelion on 12.9 Dec. 2018
UT at 1.055 au from the Sun. It made its closest ever approach to the Earth
on 16 December at only 0.078 au. It remained within 0.1 au from the Earth
for three weeks, and this provided one of the best opportunities for 
ground-based investigation of a Jupiter-family comet. 
Such an orbit makes it well suited for spacecraft exploration, and 46P was
the initial target of the ESA Rosetta mission, until it was replaced by comet
67P/Churyumov-Gerasimenko.

We observed comet 46P with the Institut de RadioAstronomie Millim\'etrique
(IRAM) 30-m telescope between 11.8 and 18.1 Dec. 2018 UT, on
21.0, 25.2 and 25.8 Dec. UT with the NOrthern Extended Millimeter Array (NOEMA)
and with the Nan\c{c}ay radio telescope. In this paper, we report
the detection of a dozen of molecules and significant upper limits for
nearly two dozen additional ones, obtained in single-dish mode.
Sect.~\ref{sect-obs} presents the observations and
Sect.~\ref{sect-data} presents the spectra of the detected molecules.
The information extracted from the observations to analyse the data and
compute production rates is provided in Sect.~\ref{sect-analysis}.
In Sect.~\ref{sect-results}, we discuss the uncertainties related to the
molecular lifetimes and present the retrieved
production rates and abundances or upper limits, which are discussed and
compared to other comets in Sect.~\ref{sect-discussion}.

\section{Observations of comet 46P/Wirtanen}
\label{sect-obs}
Comet 46P/Wirtanen was observed around the time of its closest approach to
the Sun and the Earth with several radio facilities (Nan\c{c}ay, IRAM, NOEMA,
the Atacama Large Millimeter/submillimeter Array (ALMA) and the Stratospheric
Observatory for Infrared Astronomy (SOFIA)).
We focus here on single-dish measurements obtained with the
IRAM and NOEMA millimetre radio telescopes. The OH
radical was observed with
the Nan\c{c}ay radio telescope between 1 Sept. 2018 and 28 Feb. 2019,
and observations covering the period of interest are presented in
Sect.~\ref{sect-qh2o}. ALMA data will be presented in a future paper
\citep{Biv21}, and SOFIA results are reported in \citet{Lis19}.

\setcounter{table}{0}
\begin{table*}
\caption[]{Log of millimetre observations.}\label{tablog}\vspace{-0.2cm}
\begin{center}
\begin{tabular}{rllcccc}
\hline\hline
UT date  & $<r_{h}>$  & $<\Delta>$   & Tel. & Integ. time & pwv\tablefootmark{a} & Freq. range \\
$($yyyy/mm/dd.d--dd.d) & (au)  & (au) &    & (min)\tablefootmark{b} & (mm) & (GHz)  \\
\hline
 2018/12/11.78--11.85 & 1.055 & 0.083 & IRAM &  75 & 1.7  & 248.7-256.5, 264.4-272.2 \\
         11.85--11.95 & 1.055 & 0.082 & IRAM &  81 & 3--5 & 248.7-256.5, 264.4-272.2 \\
         11.98--12.07 & 1.055 & 0.082 & IRAM &  96 & 4.2  & 240.4-248.1, 256.0-263.8 \\
         12.08--12.11 & 1.055 & 0.082 & IRAM &  27 & 4.9  & 209.7-217.5, 225.4-233.1 \\
 2018/12/12.83--12.90 & 1.055 & 0.081 & IRAM &  76 & 2--4 & 248.7-256.5, 264.4-272.2 \\
         12.91--13.01 & 1.055 & 0.080 & IRAM & 102 & 5--6 & 209.7-217.5, 225.4-233.1 \\
         13.03--13.10 & 1.055 & 0.080 & IRAM &  66 & 5--8 & 146.9-154.7, 162.6-170.4 \\
 2018/12/14.82--14.87 & 1.056 & 0.078 & IRAM &  48 & 1.4  & 248.7-256.5, 264.4-272.2 \\
         14.89--14.93 & 1.056 & 0.078 & IRAM &  42 & 2.4  & 240.4-248.1, 256.0-263.8 \\
         15.02--15.04 & 1.056 & 0.078 & IRAM &  16 & 4.3  & 240.4-248.1, 256.0-263.8 \\
         15.06--15.13 & 1.056 & 0.078 & IRAM &  64 & 4.4  & 217.8-225.6, 233.5-241.2 \\
 2018/12/15.82--15.86 & 1.056 & 0.078 & IRAM &  44 & 5.9  & 248.7-256.5, 264.4-272.2 \\
         15.88--16.02 & 1.056 & 0.078 & IRAM & 141 & 4.6  & 146.9-154.7, 162.6-170.4 \\
         16.03--16.13 & 1.056 & 0.077 & IRAM &  88 & 3.9  & 209.7-217.5, 225.4-233.1 \\
 2018/12/16.88--16.89 & 1.057 & 0.077 & IRAM &  15 & 1.1  & 248.7-256.5, 264.4-272.2 \\
         16.91--16.96 & 1.057 & 0.077 & IRAM &  53 & 1.0  & 146.9-154.7, 162.6-170.4 \\
         16.98--17.04 & 1.057 & 0.077 & IRAM &  68 & 0.4  & 209.7-217.5, 225.4-233.1 \\
         17.06--17.07 & 1.057 & 0.077 & IRAM &  19 & 0.7  & 161.3-169.1, 177.0-184.8 \\
         17.08--17.12 & 1.057 & 0.078 & IRAM &  46 & 0.8  & 240.4-248.1, 256.0-263.8 \\
 2018/12/17.81--17.83 & 1.057 & 0.078 & IRAM &  30 & 1.6  & 248.7-256.5, 264.4-272.2 \\
         17.84--17.91 & 1.057 & 0.078 & IRAM &  75 & 1.9  & 240.4-248.1, 256.0-263.8 \\
         17.98--18.09 & 1.058 & 0.078 & IRAM & 120 & 2.5  & 217.8-225.6, 233.5-241.2 \\
         18.10--18.12 & 1.058 & 0.078 & IRAM &  26 & 2.4  & 209.7-217.5, 225.4-233.1 \\
\hline
 2018/12/20.95--21.16 & 1.061 & 0.082 & NOEMA & 160 & 1--3  & 88.6+90.6 \\
 2018/12/25.15--25.25 & 1.068 & 0.094 & NOEMA &  80 & 2--3.5  & 206+208+208+211+221+223+226+227 \\
 2018/12/25.72--25.93 & 1.070 & 0.096 & NOEMA & 162 & 0.5--1  & 147+165+165.6+166+169 \\
\hline
\end{tabular}
\end{center}
\tablefoot{
  \tablefoottext{a}{Mean precipitable water vapour in the atmosphere above the telescope.}\\
  \tablefoottext{b}{Total (offset positions included) integration time (ON+OFF)
    on the source (adding 10, 8, and 9 antennas, respectively, for NOEMA).}
}
\end{table*}

\subsection{Observations with the IRAM~30-m}
\label{sect-obs-iram}
Comet 46P/Wirtanen was the target of the observing proposal 112-18
scheduled at the IRAM 30-m telescope between 11 and 18 December 2018.
Weather conditions were very good to average, with precipitable
water vapour (pwv) in the 0.4 to 8~mm range (Table~\ref{tablog}).
The worst (cloudy) conditions were at the end of the 12 Dec. run
(pwv up to 8~mm on 13.1 Dec. UT) followed by a day without observations.
Around 17.0 Dec. UT, very low opacity ($\sim$0.5~mm pwv) enabled observations
around 183~GHz.
Observations were obtained in wobbler switching mode with the secondary
(wobbling) mirror alternating pointing between the ON and 
OFF positions separated by 180\arcsec~ every 2 seconds. 

\begin{figure}
\centering
\resizebox{\hsize}{!}{\includegraphics[angle=270]{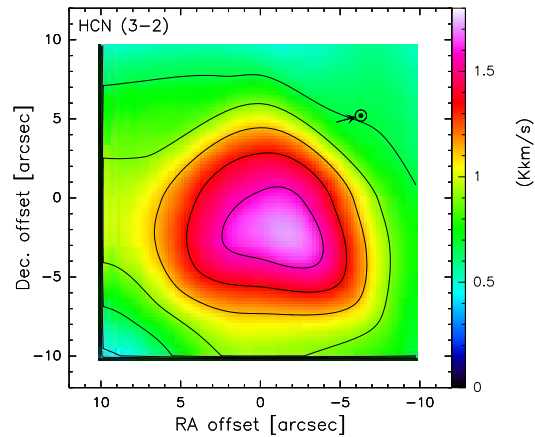}}
\caption{Map of HCN(3-2) line-integrated intensity in comet 46P/Wirtanen
  on 17.82 Dec. 2018. Offsets are relative to JPL\#181/16 position and
  corrected for estimated pointing errors. The direction of the Sun
  is indicated by the arrow in the upper right.
  Contours are drawn every 0.225~K\kms.}
\label{fighcnmap}
\end{figure}

Comet 46P was tracked using the latest JPL\#181/11
\footnote{https://ssd.jpl.nasa.gov/horizons.cgi} orbital elements,
and following orbit solutions, to compute a position in real time with
IRAM New Control System (NCS) software.
Unfortunately, the
proximity of the comet to the Earth (0.08 au) challenged the accuracy
of the NCS software and resulted in an ephemeris error of the order
of 24\arcsec~ in R.A. and 10\arcsec~ in declination, changing by a few
arc-seconds from day to day. These initial offsets were estimated
from a map of HCN(3-2) on the first day and used for the following days.
The exact ephemeris error was computed afterwards from the comparison of
the output of the NCS ephemeris and the JPL\#181/16 ephemeris solution.
The pointing was regularly checked on bright pointing
sources (rms $< 1$\arcsec).
Uranus pointing data were also used to check the beam efficiency of the antenna,
and the beam size, which follows the formula
$\theta\sim2460/\nu[{\rm GHz}]$ in \arcsec.
Coarse maps obtained on HCN (e.g. Fig.~\ref{fighcnmap}),
CH$_3$OH, and H$_2$S show an average residual
offset between the maximum brightness and the target position of
(-1\arcsec,-1\arcsec) that was used to compute the final average radial
offsets given in Table~\ref{tabobs}.

We used the EMIR \citep{Car12} 2~mm and 1~mm band receivers in 2SB mode
connected to the FTS and VESPA high-resolution spectrometers
(see Table~\ref{tablog}).

\subsection{Observations conducted with NOEMA}
\label{sect-obs-noema}
Comet 46P/Wirtanen was also observed with the NOEMA interferometer on 
21.0, 25.2, and 25.8 Dec. UT,
as part of the proposal W18AB, at 3~mm, 1~mm, and 2~mm
wavelengths. Here, we only report the ON-OFF position switching data obtained
during part of these observations for the zero-spacing information needed to
analyse interferometric data. The versatile Polyfix correlator was used to
target some dedicated molecular lines in addition to the continuum, and here we
analyse the spectra of the lines detected in this autocorrelation mode.
The frequency coverage and total integration time (adding up the 8 to
10 antennas) is provided in Table~\ref{tablog}.

\section{Spectra and line intensities}
\label{sect-data}
\begin{figure}[]
\centering
\resizebox{\hsize}{!}{\includegraphics[angle=270]{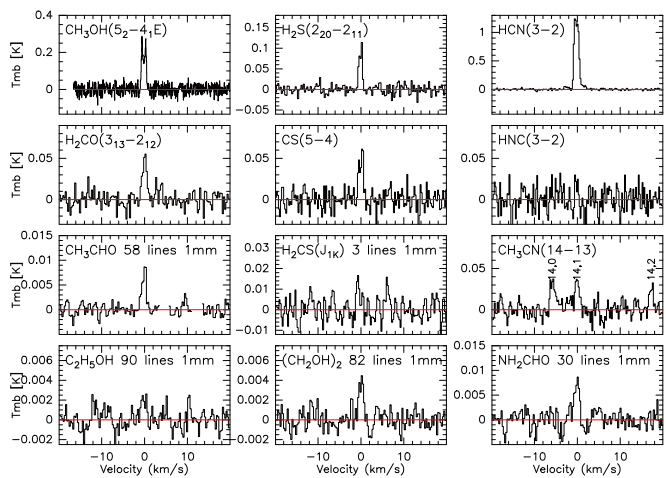}}
\caption{Molecular lines observed in comet 46P/Wirtanen between 11.9
  and 18.1 Dec. 2018. For some complex organics, we averaged several
  lines between 210 and 272~GHz (1~mm wavelength band).
The vertical scale is the main beam brightness temperature
and the horizontal scale is the Doppler velocity in the comet rest frame.}
\label{figsp1mm}
\end{figure}

\begin{figure}[]
\centering
\resizebox{\hsize}{!}{\includegraphics[angle=270]{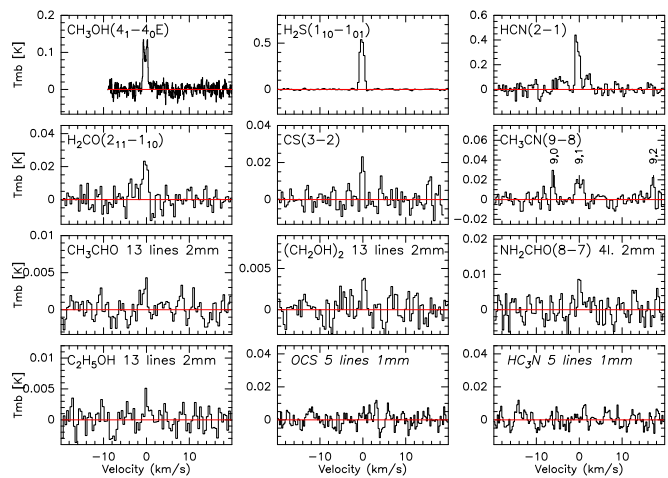}}
\caption{Molecular lines observed in comet 46P/Wirtanen between 11.9
  and 18.1 Dec. 2018 in the 2~mm band, except for HC$_3$N and OCS, for which
  we show the weighted average of five lines expected in the 1~mm band.
  For some complex organics, we averaged several
  lines between 147 and 170~GHz (2~mm wavelength).
The vertical scale is the main beam brightness temperature
and the horizontal scale is the Doppler velocity in the comet rest frame.}
\label{figsp2mm}
\end{figure}

\begin{figure}[]
\centering
\resizebox{\hsize}{!}{\includegraphics[angle=270]{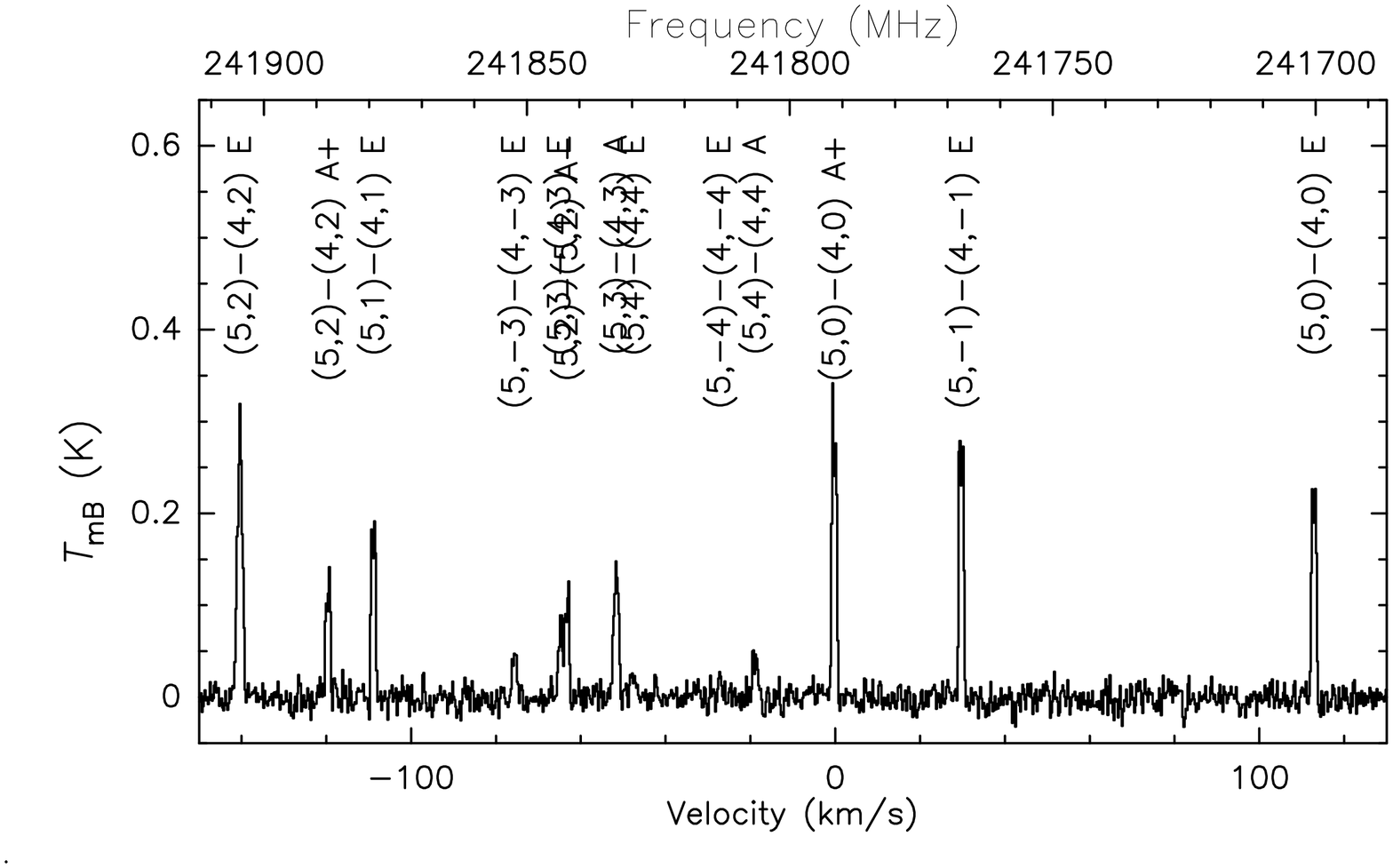}}
\caption{Series of methanol lines around 242~GHz observed in comet
  46P/Wirtanen between 12.0 and 17.9 Dec. 2018.
The vertical scale is the main beam brightness temperature,
and the horizontal scale is the Doppler velocity in the comet rest frame
and frequency in the comet frame on the upper scale.}
\label{figspmet242}
\end{figure}

\begin{figure}[]
\centering
\resizebox{\hsize}{!}{\includegraphics[angle=270]{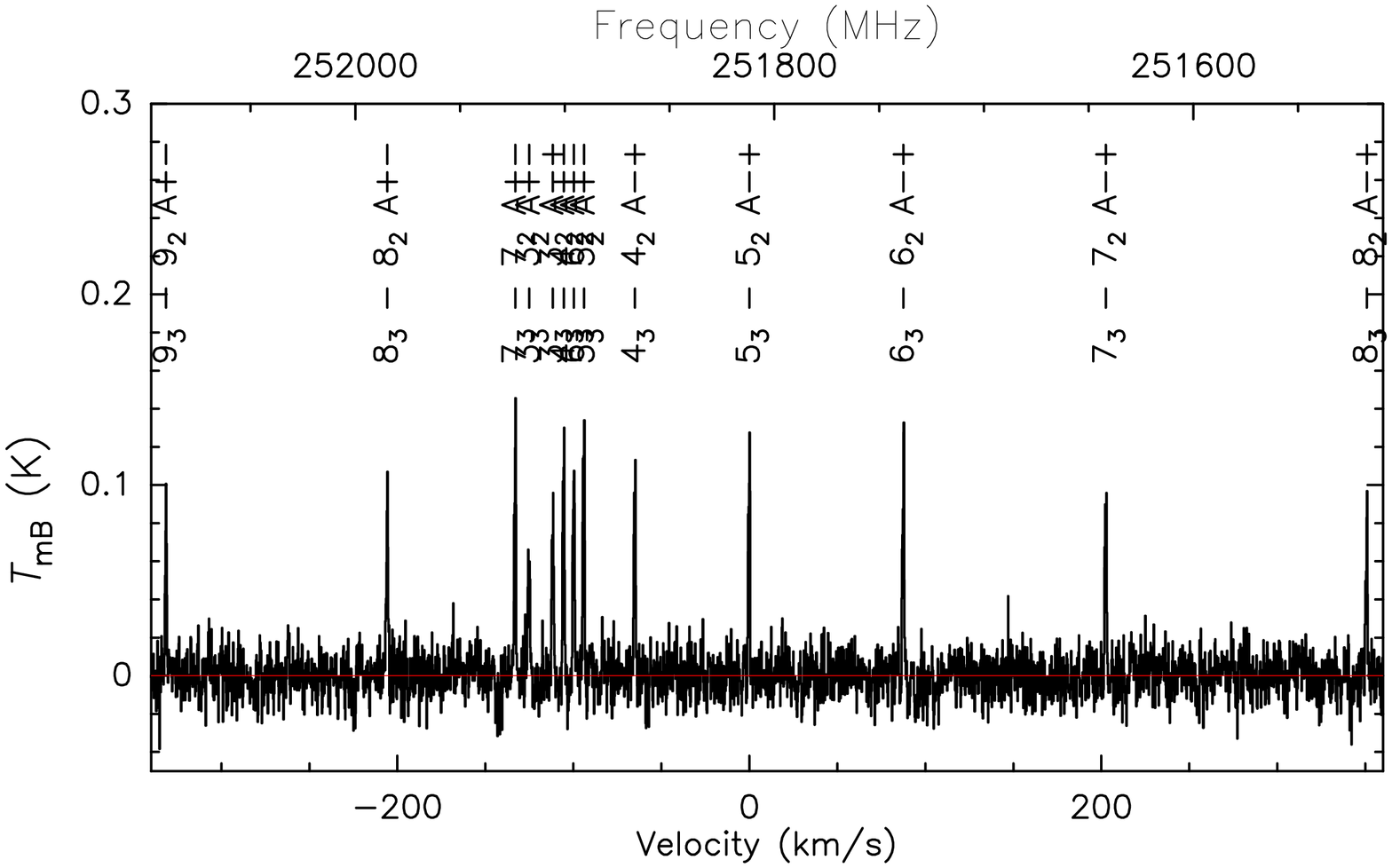}}
\caption{Series of methanol lines around 252~GHz observed in comet
  46P/Wirtanen between 11.8 and 17.8 Dec. 2018.
The vertical scale is the main beam brightness temperature,
and the horizontal scale is the Doppler velocity in the comet rest frame
and frequency in the comet frame on the upper scale.}
\label{figspmet252}
\end{figure}

A sample of IRAM spectra at 1~mm and 2~mm is shown in
Figs.~\ref{figsp1mm}--\ref{figsp2mm}. These are averages over the six nights of
observations, sampling most of the detected species in the two
  wavelength-domains.
For some species, the spectra are the weighted averages of several
lines to increase the signal-to-noise ratio (S/N),
with a weight inferred from to the noise of individual spectra.
Spectra are aligned on the velocity scale to provide
information on the line shape.
However, the computation of production rate is not based on the average line
intensity but on the weighted average of the production rates derived from each
individual line (the weight is larger for the lines expected to be the
strongest). Full spectra are shown
in the appendix (Figure~\ref{figsurvey2mm1}-\ref{figsurvey1mm6}).
Examples of CH$_3$OH lines used to derive the gas temperature are
also shown in Figs.~\ref{figspmet242} and \ref{figspmet252}.
Only spectra obtained at pointing offset $<$ 3\arcsec~ are shown.

Line intensities are given in Tables~\ref{tabobssum} and \ref{tabobs},
including values obtained at the offset position (radial averages).
Table~\ref{tabobssum} focuses on species for which several individual lines,
eventually grouped by series probing levels of similar energy, were averaged.
We also provide, for other molecules, the average intensity or upper limits
for non-detections. Where individual lines are not detected,
we selected the strongest transitions that are expected to have a
similar intensity for the averages (within a factor of 2 to 3). 

\begin{table*}
\caption[]{Line intensities from IRAM observations: multi-line averages.}\label{tabobssum}\vspace{-0.5cm}
  \renewcommand{\arraystretch}{0.87}
\begin{center}
\begin{tabular}{llllrr}
  \hline\hline
Molecule & \multicolumn{2}{c}{Transitions} & Frequency & Pointing offset & Intensity \\
         & N\tablefootmark{a} & $J_{Ka,Kc}$\tablefootmark{b} &  range [GHz] & [\arcsec] & Average [mK~\kms] \\
\hline
NH$_2$CHO & 2 &  $8_{0,8}+8_{1,8}$                 & 163--167 &  1.7 & $8.4\pm2.9$  \\
NH$_2$CHO & 3 & $10_{1,9}+10_{2,8}+10_{2,9}$       & 213--218 &  1.1 & $8.4\pm2.9$  \\
NH$_2$CHO & 5 & $10_{3,Kc}+10_{4,Kc}+10_{5,Kc}$     & 212--213 &  0.9 & $7.3\pm2.4$  \\
NH$_2$CHO & 5 & $11_{0,11}+11_{1,Kc}+11_{2,Kc}$     & 223--240 &  1.3 & $12.3\pm2.6$  \\
NH$_2$CHO & 5 & $11_{3,Kc}+11_{4,Kc}+11_{5,Kc}$     & 233--234 &  1.7 & $13.3\pm3.1$  \\
NH$_2$CHO & 5 & $12_{0,12}+12_{1,Kc}+12_{2,Kc}$     & 244--261 &  1.7 & $16.4\pm3.1$  \\
NH$_2$CHO & 5 & $12_{3,Kc}+12_{4,Kc}+12_{5,Kc}$     & 255--256 &  1.7 & $10.0\pm2.8$  \\
NH$_2$CHO & 2 & $13_{0,13}+13_{1,13}$              & 264--267 &  1.7 & $1.4\pm5.3$  \\
NH$_2$CHO & 30 & $J_{Ka,Kc}$, J=10--13, Ka=0--5    & 212--267 &  1.5 & $12.0\pm1.2$  \\
          &    &                                  &          & 10.1 & $<8.1$       \\
CH$_3$CHO & 6 &  $8_{0,8}+8_{1,8}+8_{2,7}$          & 146--154 &  1.6 & $3.4\pm1.5$  \\
CH$_3$CHO & 6 & $11_{1,10}+11_{2,9}+11_{2,10}$      & 211--217 &  0.9 & $11.4\pm2.4$  \\
CH$_3$CHO & 10 & $12_{0,12}+12_{1,Kc}+12_{2,Kc}$    & 224--236 &  1.5 &  $9.1\pm2.4$  \\
CH$_3$CHO & 8 & $12_{3,Kc}+12_{4,Kc}$               & 231--232 &  0.9 & $7.5\pm2.3$  \\
CH$_3$CHO & 10 & $13_{0,13}+13_{1,Kc}+13_{2,Kc}$     & 242--255 &  1.9 & $12.3\pm2.1$  \\
CH$_3$CHO & 8 & $13_{3,Kc}+13_{4,Kc}$               & 250--251 &  1.9 & $6.9\pm2.1$  \\
CH$_3$CHO & 6 & $13_{5,Kc}+13_{6,Kc}$               & 250.6    &  2.0 & $6.5\pm2.3$  \\
CH$_3$CHO & 6 & $14_{0,14}+14_{1,14}+14_{2,13}$     & 261--268 &  1.9 & $12.8\pm3.3$  \\
CH$_3$CHO & 4 & $14_{3,11}+14_{3,12}$               & 270--271 &  1.9 & $0.7\pm3.5$  \\
CH$_3$CHO & 32 & $J_{Ka,Kc}$, J=11--14, Ka=0--2     & 211--268 &  1.5 & $11.0\pm1.3$  \\
          &    &                                  &          & 9.5 & $9.6\pm3.0$   \\
          &    &                                  &          & 11.3 & $7.2\pm3.9$  \\
C$_2$H$_5$OH & 11 & $J_{Ka,Kc}$, J=8--10, Ka=0--1  & 147--168 &  1.9 & $5.4\pm1.9$  \\
C$_2$H$_5$OH & 26 & $J_{Ka,Kc}$, J=4--11           & 214--270 &  1.6 & $0.0\pm1.5$  \\
C$_2$H$_5$OH & 65 & $J_{Ka,Kc}$, J=12--16          & 214--267 &  1.6 & $2.9\pm0.9$  \\
(CH$_2$OH)$_2$ & 13 & $J_{Ka,Kc}$,J=12--18,Ka=0--8 & 147--169 &  1.6 & $5.2\pm1.5$  \\
(CH$_2$OH)$_2$ & 80 & $J_{Ka,Kc}$,J=20--29,Ka=0--5 & 210--271 &  1.6 & $5.0\pm0.8$  \\
          &    &                                  &          & 10.2 & $5.7\pm1.7$   \\
H$_2$CS  & 3 & $7_{17}+7_{16}+8_{18}$  & 237--271 &  1.8 & $15\pm4$  \\
\hline
HC$_3$N  & 2 & 17-16 + 18-17         & 155--164 &  1.7 & $2\pm4$  \\
HC$_3$N  & 6 & 24-23 to 29-28        & 218--264 &  1.7 & $6\pm3$  \\
OCS      & 5 & 18-17 to 22-21        & 219--267 &  1.7 & $6\pm3$  \\
HNCO     & 3 &  $10_{0,10}+11_{0,11}+12_{0,12}$ & 220--264 &  1.7 & $8\pm4$  \\
HNCO     & 9 & $J_{0,Kc}$ and $J_{1,Kc}$, J=10--12 & 219--264 &  1.7 & $1.5\pm2.2$  \\
HCOOH    & 15 & $J_{Ka,Kc}$:J=10,11:Ka<3; J=12:Ka<4 & 215--271 &  1.6 & $0.5\pm1.8$  \\
C$_2$H$_5$CN & 14 & $J_{Ka,Kc}$,J=17--20,Ka=0--6 & 147--170 &  1.6 & $1.5\pm1.6$  \\
C$_2$H$_5$CN & 61 & $J_{Ka,Kc}$,J=23--31,Ka=0--7 & 210--266 &  1.6 & $1.7\pm1.0$  \\
C$_2$H$_3$CN & 13 & $J_{Ka,Kc}$,J=16--18        & 146--171 &  1.7 & $<5.1$  \\
C$_2$H$_3$CN & 40 & $J_{Ka,Kc}$,J=22--29        & 210--272 &  1.7 & $<3.6$  \\
CH$_3$COCH$_3$ & 95 & $J_{Ka,Kc}$                & 210--272 &  1.6 & $<2.4$  \\
CH$_3$COOH   & 52 & $J_{Ka,Kc}$                & 215--271 &  1.6 & $<3.0$  \\
CH$_3$NH$_2$ & 32 & $J_{Ka,Kc}$,J=2-9,Ka=0-2   & 215--271 &  1.6 & $2.1\pm1.2$  \\
PO          & 4 & 11/2-9/2                    & 239--240 &  1.7 & $4\pm4$  \\
\hline
\hline
\end{tabular}
\end{center}
\tablefoot{
\tablefoottext{a}{Number of lines averaged}\\
\tablefoottext{b}{$J_{Ka,Kc}$ quantum numbers of the upper level of the
  transitions, with range of values, or $J_{up}-J_{low}$ for OCS, HC$_3$N, or PO.}
}
\end{table*}

\section{Data analysis}
\label{sect-analysis}
\subsection{Expansion velocity and outgassing pattern}\label{sect-vexp}
We used the line profiles with the highest S/N and those of CH$_3$OH
obtained with the high spectral resolution (40~kHz) VESPA spectrometer
(e.g. Figs~\ref{figsp1mm},\ref{figsp2mm}). As lines are often double-peaked
or asymmetric, we fitted two Gaussians, one to each of the blueshifted
($v<0)$ and redshifted ($v>0$) peaks.
The velocity of the channel at half the maximum
intensity ($VHM$) of these fits provides a good estimate of the expansion
velocity, towards the observer for the negative-velocity (blueshifted) side
and for the gas moving in the opposite direction for the positive-velocity
(redshifted) side. We found the following for the velocities at half the maximum intensity ($VHM$):
\begin{itemize}
  \item $VHM$(HCN(3-2)) = $-0.87\pm0.01$ and $+0.48\pm0.01$ \kms;
  \item $VHM$(H$_2$S($1_{10}-1_{01}$)) = $-0.86\pm0.02$ and $+0.47\pm0.02$ \kms;
  \item $VHM$(CH$_3$OH(6 lines 165-266GHz)) = $-0.83\pm0.01$ and $+0.42\pm0.02$ \kms.
\end{itemize}
On average, this suggests an expansion velocity of 0.85~\kms~ on the observer
or day side and 0.45~\kms~ on the anti-observer or night side.
For HCN, for example, if we model an hemispheric day side outgassing at
0.85~\kms~ and outgassing at 0.45~\kms~ separately in the other hemisphere,
to retrieve
the observed Doppler shift ($-0.17$~\kms) we need a production rate ratio
$Q_{day}/Q_{night}=2.3$. The corresponding total production rate
($Q_{day}+Q_{night}$) is only $2\pm2$\% higher than the value found when
assuming isotropic outgassing at a velocity of 0.65~\kms. 
Since modelling an asymmetric outgassing pattern does not significantly change
the retrieved total outgassing rates, to compute the production rates we assumed
isotropic outgassing at the mean velocity, that is, 0.65~\kms.

\subsection{Gas temperature}
\begin{figure}
\centering
\resizebox{\hsize}{!}{\includegraphics[angle=270]{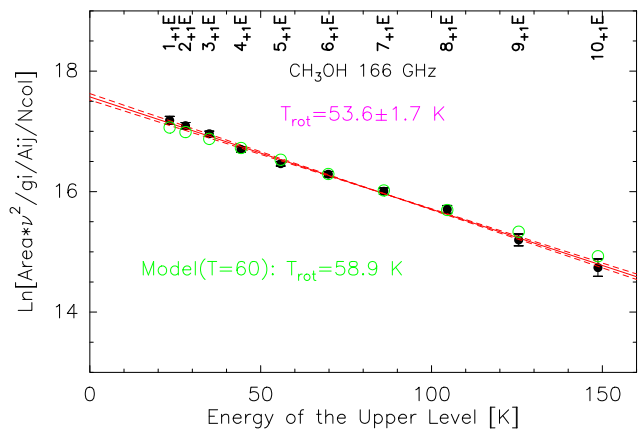}}
\caption{Rotational diagram of the 13--17 Dec. average of the methanol lines
  around 166~GHz in comet 46P/Wirtanen.
  The neperian logarithm of a quantity proportional to the line
  intensity is plotted against the energy of the upper level of each transition.
  Fit is shown by solid red line and errors are displayed by red dashed lines. 
  The black dots are the measurements and green circles the predicted values
  for a model with a gas temperature of 60 K.}
\label{figdiagrot165}
\end{figure}

\begin{figure}
\centering
\resizebox{\hsize}{!}{\includegraphics[angle=270]{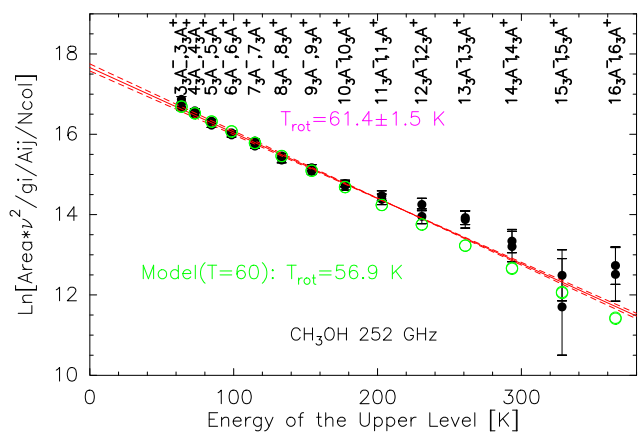}}
\caption{Same as for Fig.~\ref{figdiagrot165}, but for the 252~GHz lines of
  methanol observed between 11.8 and 17.8 Dec. UT. }
\label{figdiagrot252}
\end{figure}

\begin{figure}
\centering
\resizebox{\hsize}{!}{\includegraphics[angle=270]{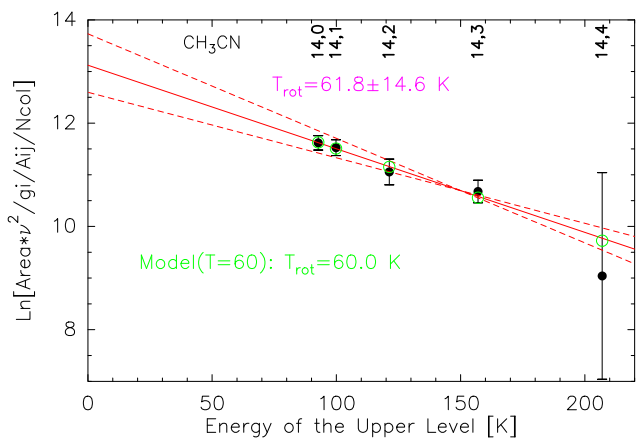}}
\caption{Same as for Fig.~\ref{figdiagrot165}, but for the 257~GHz lines of
  CH$_3$CN observed between 12.0 and 17.9 Dec. UT.}
\label{figdiagrotch3cn}
\end{figure}

\begin{figure}
\centering
\resizebox{\hsize}{!}{\includegraphics[angle=270]{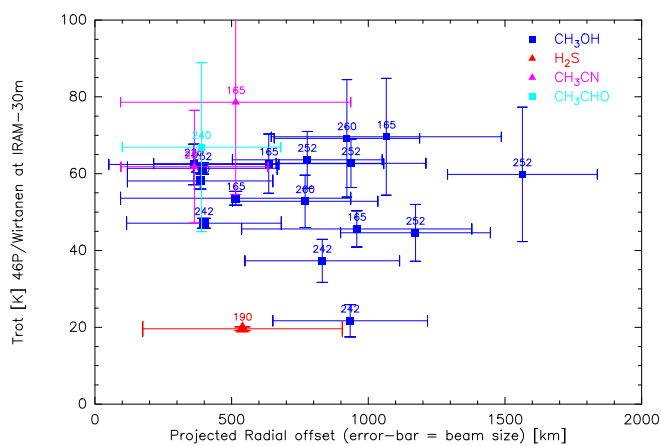}}
\caption{Plot of all rotational temperatures as a function
  of the projected pointing offset and beam size (indicating
    which region of the coma is probed).
    Most values are compatible with a
    constant gas temperature of $\approx$60 K throughout the coma.
    This includes some low rotational temperature values such as that
    derived for H$_2$S lines, of which the rotational levels are not expected to
    be thermalised.
    }
\label{figtrots}
\end{figure}

\begin{figure}
\centering
\resizebox{\hsize}{!}{\includegraphics[angle=270]{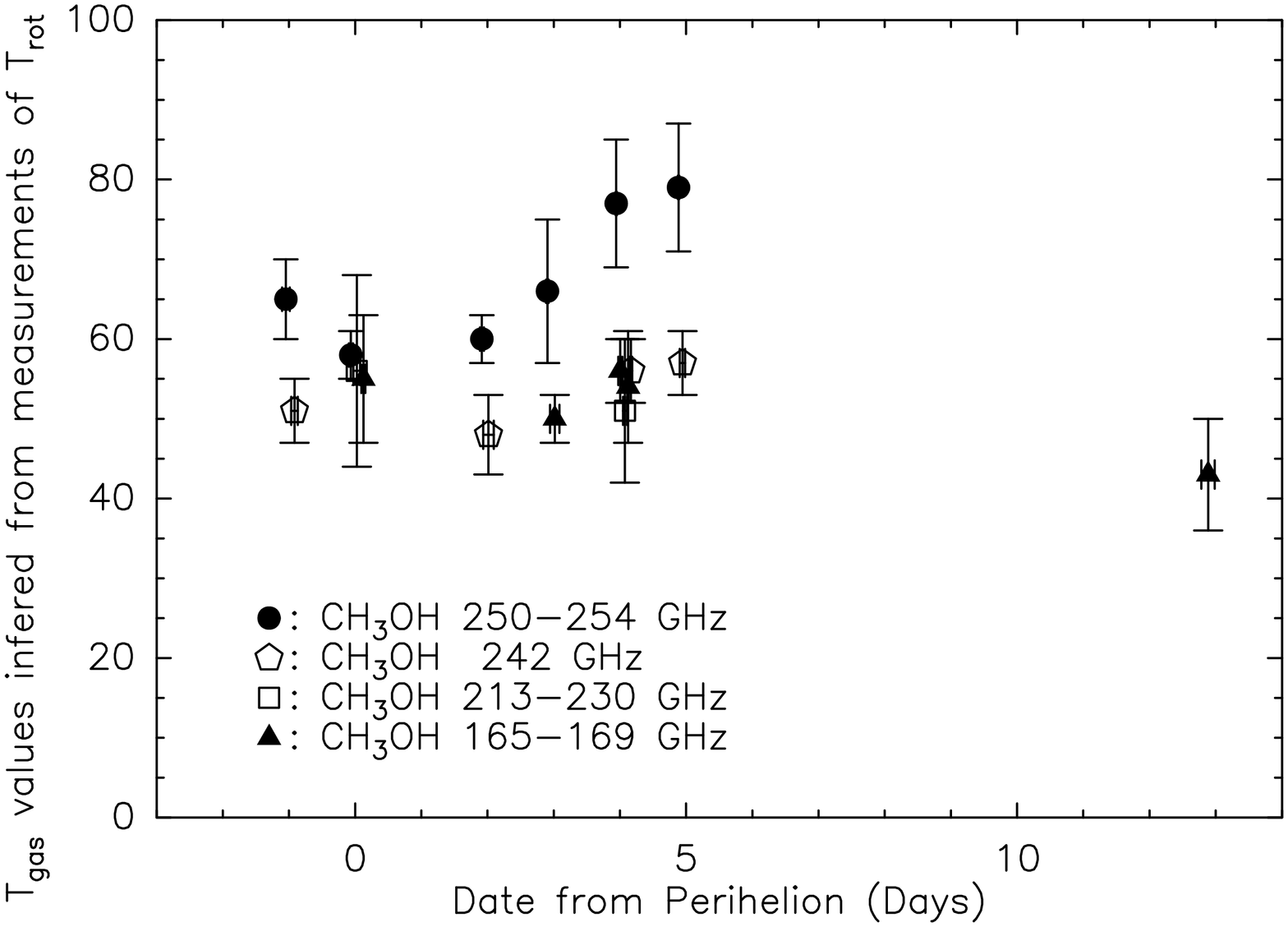}}
\caption{Plot of daily gas temperatures $T_{gas}$ derived
  from the methanol rotational lines around 167, 213-230, 242, and 252~GHz
  as presented in the first part of Table~\ref{tabtemp}.
  Variations cannot be related to the rotation of the nucleus (when
  folded on a single 9~h period). Most of the dispersion is due
  to the different series of methanol lines used: the 252~GHz ones
  sample a smaller region than the others and keep memory
  of $T_{gas}$ at larger cometocentric distances than those at 242~GHz.
  }
\label{figtgasdt}
\end{figure}

Many series of lines (especially of methanol but also CH$_3$CN) were
detected at several times during the observations. Examples of these data
as well as rotational diagrams are shown in
Figs.~\ref{figsp1mm}-\ref{figspmet252}
and Figs.~\ref{figdiagrot165}-\ref{figdiagrotch3cn}, respectively.
Inferred rotational temperatures $T_{rot}$ are given in Table~\ref{tabtemp}.
For each of these measurements, we retrieve constraints on the gas temperature,
$T_{gas}$, needed to obtain such rotational temperatures. For some series of
transitions, such as the CH$_3$OH lines at 165 or 252~GHz, $T_{gas}$ is close
and proportional to $T_{rot}$.
For CH$_3$CN, our code \citep[e.g.][]{Biv99,Biv06}
predicts $T=T_{rot}$, but for species like H$_2$S
there is a large difference due to rapid radiative decay
of the levels' population in the inner coma.
The evolution of $T_{rot}$ with nucleocentric distance 
(400--1500~km) does not show any trend (Fig.~\ref{figtrots}).
 A decrease of $T_{rot}$(CH$_3$OH 242GHz) and
a smaller one for $T_{rot}$(CH$_3$OH 252GHz) is expected and compatible
with observations. In Fig.~\ref{figtgasdt}, we also plotted
  the retrieved  $T_{gas}$ temperatures
for the daily measurements from Table~\ref{tabtemp}. The decrease of the
gas temperature with radial distance from the nucleus would
likely explain the systematically higher $T_{gas}$ deduced from the
252~GHz lines. There might also be a longer week-long trend,
but the adopted $T_{gas}=60$~K value is within $\sim2-\sigma$ of all values
in Table~\ref{tabtemp}. A $\pm10$~K variation of the gas temperature will not
impact the production rates by more than $\pm5-10$\% (using several
lines to retrieve the production rate of a molecule decreases the impact).

We also performed the rotation diagram analysis independently
for the blue-shifted and red-shifted components of the lines.
We derive the corresponding Earth ($\sim$day) side and opposite
($\sim$night) rotational temperature for the average of the
CH$_3$OH lines at 165 or 252~GHz. There is no day/night asymmetry based on the
165~GHz lines (beam$\approx$900~km), but the 252~GHz lines that probe
the gas closer ($<600$~km) to the nucleus do show a higher night-gas
temperature (Table~\ref{tabtemp}), an effect observed
in situ in the coma of comet
67P \citep{Biv19} that can be explained as a less efficient
adiabatic cooling on the night side where the outgassing rate is lower.

Infrared observations \citep{Bon20,Kha20} that probed the coma
closer to the nucleus (within 50~km) found higher gas
temperatures (80--90~K) decreasing slightly with projected distance to
the nucleus \citep[100--200~km,][]{Bon20} down to values closer to ours
(Figs.~\ref{figtrots} and \ref{figtgasdt}).
This is often the case when comparing
ground-based radio and infrared measurements, indicative of some degree
of adiabatic cooling between the distances probed by the two techniques.

\begin{table}
\renewcommand{\tabcolsep}{0.08cm}
\caption[]{Rotational temperatures and inferred gas kinetic temperatures.}\label{tabtemp}
\begin{center}
\begin{tabular}{lccrrcc}
\hline\hline\noalign{\smallskip}
UT   & Molecule & Freq. range  & lines & off.\tablefootmark{a} & $T_{rot}$\tablefootmark{b} & $T_{gas}$ \\
$($dd.d) & &  (GHz)        &\tablefootmark{c}  & (\arcsec)   & (K)  & (K) \\
\hline\noalign{\smallskip}
11.9 & CH$_3$OH & 250-254  & 22 & 1.7 & $61.7\pm4.6$  & $65\pm5$  \\
12.0 & CH$_3$OH & 241.8    & 14 & 1.7 & $44.7\pm2.4$  & $51\pm4$  \\
12.9 & CH$_3$OH & 250-254  & 22 & 1.6 & $55.1\pm3.1$  & $58\pm3$  \\
13.0 & CH$_3$OH & 213-230  &  4 & 1.0 & $63\pm13$     & $56\pm12$  \\
13.1 & CH$_3$OH & 165-169  & 10 & 1.4 & $54.2\pm8.1$  & $55\pm8$  \\
14.9 & CH$_3$OH & 250-254  & 22 & 3.1 & $56.8\pm3.1$  & $60\pm3$  \\
15.0 & CH$_3$OH & 241.8    & 14 & 1.8 & $43.1\pm3.0$  & $48\pm5$ \\
15.8 & CH$_3$OH & 250-254  & 18 & 2.0 & $62.1\pm9.2$  & $66\pm9$  \\
16.0 & CH$_3$OH & 165-169  & 10 & 1.4 & $49.4\pm2.7$  & $50\pm3$  \\
16.9 & CH$_3$OH & 250-254  & 21 & 1.1 & $72.5\pm7.9$  & $77\pm8$  \\
16.9 & CH$_3$OH & 165-169  & 10 & 2.2 & $55.0\pm3.8$  & $56\pm4$  \\
17.0 & CH$_3$OH & 213-230  &  4 & 0.7 & $57.6\pm9.0$  & $51\pm9$  \\
17.1 & CH$_3$OH & 165-169  &  9 & 1.4 & $52.7\pm6.8$  & $54\pm7$  \\
17.1 & CH$_3$OH & 241.8    & 14 & 2.0 & $48.3\pm2.6$  & $56\pm4$ \\
17.8 & CH$_3$OH & 250-254  & 22 & 1.3 & $74.0\pm6.8$  & $79\pm8$  \\
17.9 & CH$_3$OH & 241.8    & 14 & 1.2 & $50.0\pm3.0$  & $57\pm4$  \\
25.8 & CH$_3$OH & 165      &  3 & 3   & $43.2\pm7.4$  & $43\pm7$  \\
\hline
\multicolumn{6}{c}{$\approx$6-day averages} \\
\hline
16.4 & CH$_3$OH & 166     & 10 &  1.7 & $53.6\pm1.8$  & $55\pm2$  \\
     &          &         & 10 &  3.8 & $62.6\pm7.7$  & $64\pm8$  \\
     &          &         & 10 &  9.5 & $45.6\pm4.7$  & $48\pm5$  \\
     &          &         & 10 & 11.4 & $70\pm15$     & $73\pm15$ \\
16.0 & CH$_3$OH & 213-234 &  5 &  0.9 & $62.4\pm5.3$  & $58\pm5$  \\
17.6 & CH$_3$OH & 218-239 &  5 &  1.7 & $39.0\pm6.0$  & $44\pm7$  \\
16.0 & CH$_3$OH & 242     & 14 &  1.7 & $47.1\pm1.3$  & $54\pm2$ \\
     &          &         & 14 &  9.7 & $37.3\pm5.6$  & $49\pm14$ \\
     &          &         & 10 & 11.5 & $21.7\pm4.2$  & $<30$  \\
14.1 & CH$_3$OH & 252     & 28 &  2.1 & $61.4\pm1.5$  & $65\pm2$ \\
     &          &         & 22 &  8.9 & $63.6\pm7.4$  & $73\pm9$  \\
     &          &         & 20 & 11.7 & $62.7\pm6.3$  & $76\pm9$  \\
     &          &         & 20 & 15.9 & $44.6\pm7.4$  & $55\pm11$  \\
     &          &         & 18 & 22.8 & $60\pm17$     & $90\pm40$ \\
14.1 & CH$_3$OH & 250-267 &  5 &  2.1 & $58.1\pm2.1$  & $59\pm2$ \\
     &          &         &  5 &  8.9 & $52.8\pm6.8$  & $55\pm7$  \\
16.4 & CH$_3$CN & 165.4   &  5 &  1.7 & $79^{+30}_{-19}$ & $60-110$  \\
16.0 & CH$_3$CN & 257.3   &  5 &  1.7 & $62\pm15$     & $47-77$  \\
16.2 & H$_2$S   & 169,216 &  2 &   2  & $19.6\pm0.5$  & $\sim50$  \\
15.0 & CH$_3$CHO &149-270 & 68 & 1--2 & $67^{+29}_{-15}$ & $52-96$ \\ 
\hline
\multicolumn{6}{c}{$\approx$6-days averages: day side (based on line Area($v<0$)):} \\
\hline
12-18 & CH$_3$OH & 252  & 28 & 2.1 & $57.2\pm1.9$ & $60\pm3$ \\
12-18 & CH$_3$OH & 165  & 10 & 1.7 & $53.0\pm2.1$ & $54\pm2$ \\
\hline
\multicolumn{6}{c}{$\approx$6-days averages: night side (based on line Area($v>0$)):} \\
\hline
12-18 & CH$_3$OH & 252  & 27 & 2.1 & $71.0\pm2.5$ & $75\pm4$ \\
12-18 & CH$_3$OH & 165  & 10 & 1.7 & $52.5\pm3.0$ & $53\pm3$ \\
\hline
\end{tabular}
\end{center}
\tablefoot{
  \tablefoottext{a}{Mean pointing offset.}\\
  \tablefoottext{b}{Result of non-linear fit with $\chi^2$ minimisation.}\\
  \tablefoottext{c}{Number of lines used for the determination of $T_{rot}$.}
}
\end{table}

\subsection{Reference water production rate}\label{sect-qh2o}
We used the water production rates derived from the
observation of the H$_2^{18}$O line with the SOFIA airborne
observatory -- assuming $^{16}$O/$^{18}$O = 500 \citep{Lis19}.
These observations were obtained with a somewhat larger beam
(50\arcsec; although smaller than the Nan\c{c}ay or SWAN fields of
view) but cover the same period (14--20 Dec. 2018)
and were analysed with the same numerical codes. The average value
of $Q_{\rm H_2O}=0.8\times10^{28}$ \mols~ was used for the computation of
excitation conditions (collisions) and relative abundances.
Water production rates measured from infrared observations at the same
time \citep{Bon20} yield about the same value.
The measurements of \citet{Com20} of Lyman-$\alpha$ emission with the
SOHO/SWAN experiment do not cover
the period of closest approach to the Earth, but derived
$Q_{\rm H_2O}=0.8\times10^{28}$ \mols~ on 10.9 Dec. 2018.

We also tried to detect directly the H$_2$O($3_{13}-2_{20}$) line at
183.3~GHz (Fig.~\ref{figsurvey2mm3}) during a period of very good
weather at IRAM, but we could not integrate long enough and the
weather proved too marginal for this observation.
In addition the calibration uncertainty is likely large at this 
frequency ($\pm50$\% probably): on the one hand the HCN(2-1) line in the
same band and a methanol line yield a correct production rate, but on the
other hand the water maser line observed in Orion spectra acquired
for calibration purposes is somewhat stronger than expected.
We obtain a $3-\sigma$ upper limit 
$Q_{\rm H_2O}< 5\times10^{28}$ \mols,~ which is a factor $\sim6$ times
higher than the other measured values.
This has been derived using the same approach as for all other
molecules as detailed in the next section.
Line opacity is taken into account both for excitation
  \citep[as detailed in e.g.][]{Biv07} and radiative transfer.\\

\subsection{Production rates of OH from Nan\c{c}ay observations}

The OH 18-cm lines were observed in 46P with the Nan\c{c}ay radio telescope
from 12 to 20 December, thanks to a small positive inversion of the
OH maser.  The OH inversion peaked on 16 December when the heliocentric
radial velocity was 1.0~\kms, at 0.040 according to \citet{Des81}
or at 0.099 based on \citet{Sch88}.

\begin{figure}
\centering
\resizebox{\hsize}{!}{\includegraphics[angle=270]{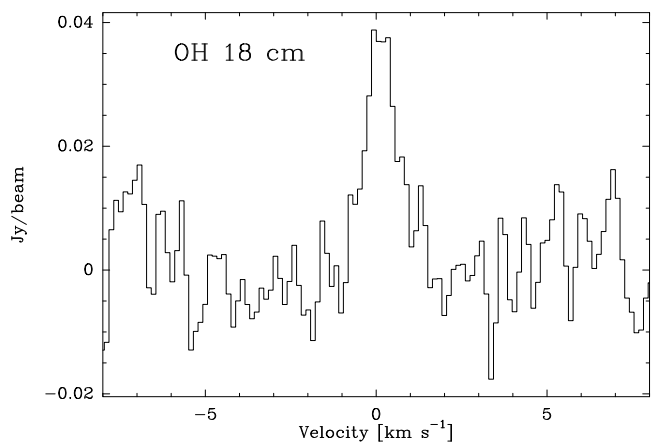}}
\caption{Eighteen-centimetre OH line of comet 46P/Wirtanen observed with the
Nan\c{c}ay radio telescope from 12 to 20 Dec. 2018.  The 1665 and 1667
MHz lines, both circular polarisations scaled to the 1667 MHz line, have
been averaged.}
\label{figoh}
\end{figure}

The time-averaged spectrum is shown in Fig.~\ref{figoh}. An emission line is
definitely detected.  The corresponding OH production rate is
$1.7\pm0.4\times10^{28}$\mols~
using our nominal model with quenching \citep[see][]{Cro02} and the
inversion of Despois et al.  For the inversion of Schleicher \& A'Hearn, the
production rate is lowered to $0.7\pm0.2\times10^{28}$\mols.
These values, although
imprecise due to the uncertainty on the OH excitation, are in line with the
other measurements of the water production rate.

The OH line is observed to be unusually narrow with a FWHM of $1.1\pm0.2$~\kms.
This is comparable to the line widths observed for parent
molecules (Sect.~\ref{sect-vexp}) and may be due to the thermalisation of the
OH radical in the near-nucleus region.  This may also be due to the Greenstein
effect, which is the differential Swings effect within the coma \citep{Gre58}.
  For the geometry of the
observation and the OH inversion as a function of the heliocentric radial
velocity, the expected OH inversion significantly drops for OH radicals at,
for example, $\pm1$~\kms~ radial velocity.

\section{Production rates and abundances}
\label{sect-results}

We determine the production rate or an upper limit for 35 molecules.
We assumed that all species follow a parent or daughter Haser
  density distribution with a destruction scale length, as discussed
  in the next section and listed in Table~\ref{tabmolparam}.
  CS and SO are assumed to
come from the photo-dissociation of CS$_2$ and SO$_2,$ respectively.
H$_2$CO, as specified in Table~\ref{tabqpmoy}, is assumed to come
from a distributed source with a scale length of 5000~km \citep{Biv99},
which fits observations obtained at various offsets and with different beam
sizes (lines observed at 2~mm and 1~mm wavelengths).
All production rates are provided in Tables~\ref{tabqp} and
\ref{tabqpmoy}.

\subsection{Molecular lifetimes}
\label{sect-lifetime}
One major source of uncertainty for the abundance of several (complex)
molecules is their photo-destruction rate (dissociation and ionisation).
For a molecule with a short lifetime, the uncertainty in
the retrieved production rate can be as large as the uncertainty in the
photo-destruction rate, but thanks
to the proximity of comet 46P/Wirtanen to the Earth (0.08 au), the small
beam size of IRAM mostly probed `young' molecules: 90\% of the signal
comes from molecules younger than $\approx2.5\times10^3$~s, which
should reduce the impact of the uncertainty in their photolytic lifetime.
In addition, the solar activity was close to a deep minimum, which
caused the photo-dissociation rates to be close to their minimum values
and reduced other effects such as, for example, dissociative impact from ions.

Table~\ref{tabmolparam} summarises the latest information available on the
lifetime of these molecules and the values that we adopted. For some
well-observed molecules, we applied a solar-activity-dependent correction
based on the 10.7-cm solar flux unit (70 sfu on average for the period,
mostly representative of the `quiet' Sun) as in \citet{Cro89}.
For several molecules, there is a large range of published values for their
photo-destruction rate, or no values at all. For detected molecules, we can also
obtain some constraints from their line shape. Since the expansion velocity
generally increases with distance from the nucleus (e.g. acceleration due to
photolytic heating), molecules with longer lifetimes will likely exhibit broader
lines.
This was clearly evidenced for comets at small heliocentric distances
\citep{Biv11}. Table~\ref{tabmolparam} and details hereafter provide
some loose constraints we can derive for some molecules. Observational data used
for this purpose are presented in Appendix~\ref{app-linewidth}. 
The H$_2$S, HCN, and CH$_3$OH molecules have well established photo-dissociation
rates, which can be used as references. We note that radiative excitation can
also affect the line width, for example if the upper rotational state of
the observed transition is depopulated at a faster rate than it is modelled.
However, in that case a bias in the lifetime compensates for the weakness of the
excitation model. For secondary species such as CS, SO, and H$_2$CO, on the
other hand, any excess energy converted into velocity might affect the
line shape, especially outside the collision-dominated region.


  {\bf HC$_3$N}: in \citet{Biv11}, cyanoacetylene clearly showed a
  narrower line than HCN, CH$_3$OH, and CS because of its shorter lifetime.
  It was detected with a high resolution and high S/N in the comet
  Hale-Bopp in April-May 1997 \citep{Boc00}, but its
  line shape did not suggest a much shorter lifetime than HCN or CH$_3$OH,
  while the mean width of HC$_3$N line is closer to that of H$_2$S in
  comet C/2014~Q2 \citep[][and Table~\ref{tablinewidth}]{Biv15}.
  Thus, the adopted photo-dissociation rate is in agreement with the range
  suggested by those observations.
  
  {\bf CH$_3$CN}: methyl cyanide has a relatively long
  lifetime according to \citet{Cro94} and \citet{Hea17}, but most
  observations, including those of 46P
  (Figure~\ref{figsp2mm},\ref{figsp1mm}, Table~\ref{tablinewidth}),
  show lines with widths similar to or slightly smaller than
  those of HCN or methanol. This suggests that its
  photo-destruction rate might be a few times larger than assumed, although
  this does not affect the retrieved production rate in comet 46P. In addition
  the rotational and derived kinetic temperature of CH$_3$CN is most often
  comparable to or higher than the one derived from CH$_3$OH
  (Fig.~\ref{figtrots}), again suggesting that the CH$_3$CN lines might probe
  the inner coma more, which is compatible with a shorter lifetime.
  Assuming $\beta_0=2\times10^{-5}$s$^{-1}$
  for the observations of comet 46P will not increase the production rate by
  more than 2\%, which is negligible.
  
  {\bf NH$_2$CHO:} the lifetime of formamide is poorly known,
  but constraints from observations already discussed in \citet{Biv14}
  and from line widths do suggest that its photo-destruction rate from
  \citet{Hea17}, similarly to the value we adopted, is a good estimate.
  
  {\bf CO:} the lifetime of carbon monoxide is very long, but the CO line
  never shows a width larger than the species with a photo-dissociation rate
  around $\beta_0=10^{-5}$s$^{-1}$. Rather than an underestimation of its
  photo-ionisation or dissociation rate, this might be the result of,
  for example, a low
  translational collisional cross-section, which would make CO less sensitive
  to gas acceleration by photolytic heating, given the small size and dipole
  moment of the molecule.
  
  {\bf CH$_3$CHO:} acetaldehyde has now been detected in six
  comets, and in all cases the width of the CH$_3$CHO lines is smaller
  than those of HCN and CH$_3$OH but slightly larger than that of H$_2$S.
  \citet{Hue15} and \citet{Hea17} suggested higher photo-destruction rates
  ($\beta_0=18\times10^{-5}$s$^{-1}$) than the one we have
  used so far \citep{Cro04}, but it is compatible with the observations.
  This might increase the retrieved abundance in comets observed further away
  from the Earth, but for comet 46P it only increases it by $\sim8$\%.
  However, in the case of comet C/2013~R1 \citep{Biv14}, if the lifetime of
  CH$_3$CHO is shorter than assumed, this might also significantly increase its
  abundance as it gets closer to the Sun, and this would need to be explained.
  
  {\bf HCOOH:} there are few observations of formic acid with a
  good S/N (mostly in Hale-Bopp and C/2014~Q2), but they all suggest
  that HCOOH does not expand much faster
  than H$_2$S, so its photo-destruction rate might be closer to
  $\beta_0=10\times10^{-5}$s$^{-1}$, but it is likely not as large as was
  suggested by \citet{Hue15}. Using this value or the
  one of \citet{Cro94} does not affect the abundance determination
  in 46P, but if we use $\beta_0=90\times10^{-5}$s$^{-1}$ instead, as suggested
  by \citet{Hue15}, the limit on the abundance in 46P is 40\% higher,
  and likely to be much higher in other comets.
  
  {\bf (CH$_2$OH)$_2$:} ethylene glycol has an unknown lifetime,
  but since the first identification of this molecule in a comet \citep{Cro04},
  we used $\beta_0=2\times10^{-5}$s$^{-1}$. There are few
  high spectral resolution detections with a good S/N. \citet{Biv14,Biv15} and
  \citet{Biv19}
  show the average of several lines in a few comets: the line shape is
  generally narrower than it is for HCN and CH$_3$OH, but the uncertainty
  is too large for definite conclusions -- except in C/2014~Q2, where it seems
  clearly narrower than for HCN lines. A lifetime a few times shorter than
  assumed is possible. As for HCOOH, an underestimate of the photo-destruction
  rate could explain the apparent decrease in abundance at lower heliocentric
  distances observed in comet C/2013~R1 \citep{Biv14}.
  Using $\beta_0=5\times10^{-5}$s$^{-1}$ for comet 46P will not change the
  retrieved production rate.
  
  {\bf H$_2$CS:} thioformaldehyde was first detected through a
  single line in the comet Hale-Bopp \citep{Woo99}. Since that time,
  it has only been observed with sufficient spectral resolution
  and S/N in the comet C/2014~Q2 \citep{Biv15}, although this is insufficient
  to accurately measure the line width. The line width seems barely
  larger than the H$_2$S one. Therefore, in the absence of any photolysis
  information, a lifetime comparable to or a bit longer than H$_2$S looks like a
  reasonable assumption.
  
  {\bf CH$_3$NH$_2$:} methylamine has not yet been detected
  remotely in comets. We assumed $\beta_0=2\times10^{-4}$s$^{-1}$
  in \citet{Biv19}, which is comparable to the value from \citet{Hea17}
  but lower than the one referred to in \citet{Cro94}.
  
  {\bf CH$_3$SH:} methanethiol (or methyl mercaptan) has only
  been detected in situ in comet 67P with an abundance relative to water
  of 0.04\% \citep{Rub19}. In \citet{Cro04}, we assumed a photo-dissociation
  rate $\beta_0=10^{-4}$s$^{-1}$, whereas recent evaluations give
  values 25$\times$ or 3$\times$ higher \citep{Hue15,Hea17}, 
  with some confidence. With such short lifetimes, the assumed
  value of $\beta_0$(CH$_3$SH) will directly impact the retrieved abundance.
  We will use $\beta_0=5\times10^{-4}$s$^{-1}$. With the \citet{Hue15} value,
  which is $5\times$ higher, the retrieved production rate (upper limit)
  would be a factor of 2 higher.

For other molecules, either the available photo-destruction rates are
compatible with observations, or no information at all is available
and we must rely on a
priori values $\beta_0=2-10\times10^{-4}$s$^{-1}$ as used in some previous
studies \citep[e.g.][]{Cro04,Cro04b} generally based on similar molecules.
Figure~\ref{figcolbeta0} can be used to infer the change of column density and
inversely the production rate, due to a change in the photo-destruction rate.

\begin{table}
 \renewcommand{\tabcolsep}{0.00cm}
\caption[]{Photo-dissociation + photo-ionisation rates of the molecules at $r_h$ = 1.0 au.}\label{tabmolparam}
\begin{center}
\begin{tabular}{lccccc}
\hline\hline\noalign{\smallskip}
Molecule  & \multicolumn{5}{c}{Photo-dissociation+photo-ionisation rate [$\times10^{-5}$s$^{-1}$]} \\
&  C1994\tablefootmark{a} & H2015\tablefootmark{b} & H2017\tablefootmark{c}
                                   & Adopted\tablefootmark{d}     & Comets\tablefootmark{e} \\
\hline
H$_2$O    & 1.3   &  1.21--2.20   &  {\bf 0.9}  &  1.21  &  \\
HCN       & 1.5   &  1.31--3.24   &  {\bf 0.8}  &  1.54  &  \\
HNC       &       &      --       &             & {\sl 1.54}  &  \\
HC$_3$N   & 6.6   &  3.92--6.79   &  3.7        &  6.6   &  2--10 \\
CH$_3$CN  & 0.67  &  1.12--2.63   &  {\bf 0.47} &  0.68  &  1--5  \\
HNCO      & 2.9   &  2.87--5.08   &             &  2.9   &  2--10 \\
NH$_2$CHO & 67    &      --       &  {\bf 10.3} &  10.0  &  $\sim10$\\
CH$_3$NH$_2$ & 67 &      --       &  {\bf 24.4} &  20.0  &  \\
C$_2$H$_3$CN &    &      --       &             & {\sl 2}  &  \\
C$_2$H$_5$CN &    &      --       &             & {\sl 2}  &  \\

CH$_3$OH  & 1.3   &  1.14-2.07    &  {\bf 0.85} &  1.31  &  \\
H$_2$CO   & 20    &   21--22      &      19.3   &  20.0  &  1--8\tablefootmark{f}\\
CO        & 0.075 & 0.075--0.188  &  {\bf 0.055}&  0.065 &  $<0.5$--2 \\
CH$_3$CHO & 7.5   & 17.9--18.2    &  {\bf 18.2} &  18.0  & 8--20 \\
HCOOH     & 3.2   &   88--113     &             &  10.0  & 5--15 \\
C$_2$H$_5$OH & 1.8 &    --        &  {\bf 1.5}  &   1.8  & 2--5 \\
(CH$_2$OH)$_2$ &  &     --        &             &   2--5  & 3--10 \\
HCOOCH$_3$ & 4.7  &     --        &             &   4.7  &  \\
CH$_2$OHCHO &     &     --        &             &   {\sl 2}  &  \\
CH$_2$CO   & 44   &     --        &             &   44  &  \\
CH$_3$COOH & 5.1  &     --        &             &   5.1  &  \\
CH$_3$OCH$_3$ & 3.1 &   --        &             &   3.1  &  \\
CH$_3$COCH$_3$ &  &     --        &             &   5.0  &  \\
c-C$_2$H$_4$O &   &     --        &             &  {\sl 10}  &  \\

H$_2$S     & 25   &  32.6--26.8   &  {\bf 23.1} &   25  &  \\
CS         &      &     --        &  {\sl 1.2}  &  2.5\tablefootmark{g}  &  \\
H$_2$CS    &      &     --        &             &   20\tablefootmark{h}  &  2--30 \\
OCS        & 9.4  &  10.2--13.0   &  {\bf 6.9}  &  9.4  &  4--20\\
SO         & 15   &    62--66     &  {\sl 45 }  &   15  &  $\sim$10\tablefootmark{f} \\
SO$_2$     & 21   &    25--28     &  {\bf 19.3} &   25  &  \\
CH$_3$SH   &      &   250--253    &     32.6    &   50  &  \\
NS         &      &     --        &             &    5\tablefootmark{i}  &  \\

Glycine    &      &     --        &             &  1000\tablefootmark{j} & \\
PH$_3$     &      &   6.1--7.6    &             &  6.1  &  \\
PO         &      &     --        &             &  {\sl 10}   &  \\
PN         &      &     --        &             &  {\sl 10}   &  \\
\hline
\end{tabular}
\end{center}
\tablefoot{
\tablefoottext{a}{From \citet{Cro94} and references therein.}\\
\tablefoottext{b}{For the quiet and active Sun from \citet{Hue15}.}\\
\tablefoottext{c}{Solar value from \citet{Hea17}:
  values given in bold have an uncertainty below 30\%,
  and values in italics have an uncertainty larger than a factor of 2.}\\
\tablefootmark{d}{Assumed values with no constraint are in italics.}\\
\tablefootmark{e}{Constraints obtained from line width
(see text, Appendix~\ref{app-linewidth}, and e.g. \citet{Biv11}).}\\
\tablefootmark{f}{For H$_2$CO and SO, the constraint from line width is on
  the combined scale length of parent + daughter. For SO, it is compatible
  with the assumed SO$_2$ and SO lifetimes (see also \citet{Boi07}).}\\
\tablefootmark{g}{Estimate from \citet{Biv11}.}\\
\tablefootmark{h}{Value proposed by \citet{Woo99}.}\\
\tablefootmark{i}{
  Centre of the range proposed by \citet{Irv00}, which is estimated
  from the analogy between NS and NO and the fact that sulphur species
  have shorter lifetimes.}\\
\tablefootmark{j}{Value used in \citet{Had19}.}\\
}
\end{table}

\subsection{Evolution of production rates}
\label{sect-qps}
\begin{figure}
\centering
\resizebox{\hsize}{!}{\includegraphics[angle=0]{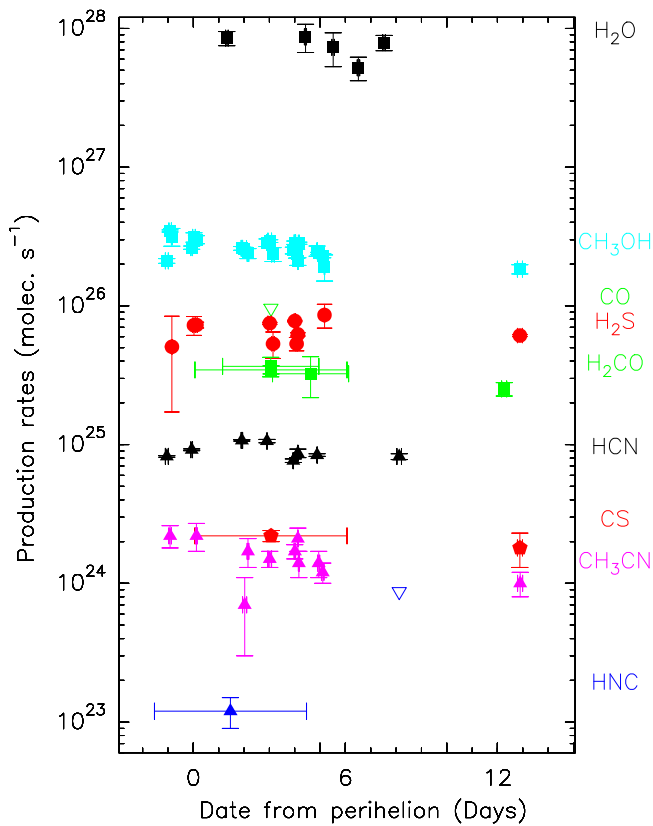}}
\caption{Evolution of production rates in comet 46P/Wirtanen
  between 12 and 25 Dec. 2019 (perihelion was on 12.9 Dec.) for
  the nine main molecules. Downward-pointing empty triangles are
  3-$\sigma$ upper limits. Coloured symbols are in the vertical
  order of the corresponding molecules' names on the right.
  Water production rates have been computed from data in \citet{Lis19}.}
\label{figqpdt}
\end{figure}

As we observed several molecules regularly (such as HCN, which was
mapped daily to check the pointing, or CH$_3$OH, which shows lines in every
observing setup), we also looked for daily variations of the activity.
Table~\ref{tabqp} provides the derived production rates for each time interval,
corresponding to 1--2~h integration on a specific observing setup.
The values are derived from both the on-nucleus pointing and
offset positions when the S/Ns are high enough (e.g. HCN).
Production rates are given for the four molecules (HCN, CH$_3$OH,
CH$_3$CN, and H$_2$S) that are detected with
sufficient S/Ns at different epochs.
Production rates as a function of time are
shown in Fig.~\ref{figqpdt}. Time
variations do not exceed
$\pm20-30$\% from the average value.
An apparent rotation period of $\approx$9~h was reported
\citep{Han19,Mou19,Far21},
and we tried to fold the production rates
or line Doppler shift values on periods
of 4 to 18~h, but no clear pattern appears for a specific period.

\begin{table*}
\caption[]{Daily production rates in comet 46P in December 2018.}\label{tabqp}
\begin{center}
\begin{tabular}{lccccc}
\hline\hline\noalign{\smallskip}
UT day  & $r_h$  & \multicolumn{4}{c}{Production rate $Q$ [$\times10^{25}$\mols]} \\
$[$dd.dd] &  [au]  & HCN      & CH$_3$OH     & H$_2$S     & CH$_3$CN \\
\hline
11.89 & 1.055 & $0.82\pm0.01$ & $21.1\pm0.8$ & & \\
12.02 & 1.055 &               & $34.7\pm0.6$ &             & $0.22\pm0.04$\\
12.09 & 1.055 &               & $31.5\pm4.6$ & $5.1\pm3.3$ & \\
12.87 & 1.055 & $0.92\pm0.01$ & $26.3\pm0.7$ & & \\
12.96 & 1.055 &               & $31.2\pm2.4$ & $7.2\pm1.1$ & \\
13.06 & 1.055 &               & $30.0\pm2.0$ & $7.3\pm0.4$ & $0.22\pm0.05$\\
14.85 & 1.056 & $1.07\pm0.01$ & $26.0\pm0.8$ & & \\
14.95 & 1.056 &               & $25.8\pm0.6$ &             & $0.07\pm0.04$\\
15.09 & 1.056 &               & $23.9\pm5.0$ &             & $0.17\pm0.04$\\
15.84 & 1.056 & $1.06\pm0.03$ & $28.6\pm1.8$ & & \\
15.95 & 1.056 &               & $29.3\pm0.7$ & $7.5\pm0.1$ & $0.15\pm0.02$\\
16.08 & 1.056 &               & $23.6\pm2.7$ & $5.3\pm1.2$ & \\
16.88 & 1.057 & $0.77\pm0.02$ & $25.1\pm1.4$ & & \\
16.94 & 1.057 &               & $28.5\pm0.8$ & $7.8\pm0.1$ & $0.17\pm0.02$\\
17.01 & 1.057 &               & $23.8\pm1.7$ & $5.3\pm0.6$ & \\
17.06 & 1.057 & $0.87\pm0.06$ & $20.9\pm1.3$ & $6.2\pm0.2$ & $0.21\pm0.04$\\
17.10 & 1.057 &               & $28.3\pm0.5$ &             & $0.14\pm0.03$\\
17.82 & 1.057 & $0.84\pm0.02$ & $24.9\pm2.1$ & & \\
17.88 & 1.057 &               & $23.9\pm0.4$ &             & $0.14\pm0.03$\\
18.04 & 1.058 &               & $23.0\pm0.8$ &             & $0.12\pm0.02$\\
18.11 & 1.058 &               & $19.2\pm4.1$ & $8.6\pm1.7$ & \\
21.06 & 1.061 & $0.82\pm0.04$ &              & & \\
25.83 & 1.070 &               & $18.4\pm1.4$ & $6.1\pm0.1$ & $0.10\pm0.02$\\
\hline
\end{tabular}
\end{center}
\end{table*}

\begin{table}\renewcommand{\tabcolsep}{0.07cm}
\caption[]{Average production rates in 46P in December 2018.}\label{tabqpmoy}
\begin{center}
\begin{tabular}{lccccc}
\hline\hline\noalign{\smallskip}
UT date  & Molecule & $r_h$  & Production rate $Q$ & Lines\tablefootmark{a} \\
$[$dd.d-dd.d] &      &  [au]  & [$\times10^{25}$\mols] &  \\
\hline
11.8--17.8 & HCN        & 1.056  & $0.92\pm0.01$  &  2 \\
11.8--17.8 & HNC        & 1.056  & $0.011\pm0.003$ &  1 \\
           & HNCext\tablefootmark{b} & & $0.030\pm0.009$ &  1 \\
12.0--18.1 & CH$_3$CN   & 1.056  & $0.137\pm0.007$ & 20 \\
11.8--18.1 & CH$_3$OH   & 1.056  & $26.5\pm0.2$   & 71 \\
12.1--18.1 & H$_2$S     & 1.056  & $7.31\pm0.07$ & 2 \\
13.1--17.1 & H$_2^{34}$S & 1.056  & $0.36\pm0.05$ & 1 \\
12.0--17.9 & CS         & 1.056  & $0.22\pm0.02$ & 2 \\
25.83      & CS         & 1.070  & $0.18\pm0.05$ & 1 \\
12.1--18.1 & H$_2$CO    & 1.056  & $0.37\pm0.08$ & 4 \\
           & H$_2$COext\tablefootmark{c} &       & $3.5\pm0.4$ & 4 \\
25.20      & H$_2$CO    & 1.068  & $0.47\pm0.05$ & 2 \\
           & H$_2$COext\tablefootmark{c} &       & $2.5\pm0.3$ & 2 \\
11.9--18.1 & NH$_2$CHO  & 1.056  & $0.12\pm0.01$ & (32) \\
11.9--18.1 & CH$_3$CHO  & 1.056  & $0.50\pm0.05$ & (72) \\
11.9--18.1 & C$_2$H$_5$OH & 1.056 & $0.90\pm0.28$ & (97) \\
11.9--18.1 & (CH$_2$OH)$_2$& 1.056 & $1.5\pm0.2$ & (93) \\
12.1--18.1 & CO         & 1.056  & $<9.8$ & (1) \\
13.1--17.1 & H$_2^{33}$S & 1.056  & $<0.36$ & (1) \\
12.0--18.1 & HDS        & 1.056  & $<0.43$ & (2) \\
11.8--18.1 & OCS        & 1.056  & $<0.56$ & (6) \\
11.8--18.1 & H$_2$CS    & 1.056  & $\sim0.13$ & (7) \\
12.0--18.1 & SO         & 1.056  & $<0.72$ & (5) \\
11.8--18.1 & SO$_2$     & 1.056  & $<0.24$ & (5) \\
12.0--17.9 & NS         & 1.056  & $<0.09$ & (2) \\
11.9--18.1 & CH$_3$SH   & 1.056  & $<0.49$ & (30) \\
12.0--18.1 & H$^{13}$CN & 1.056  & $<0.014$ & (1) \\
12.0--18.1 & HC$^{15}$N & 1.056  & $<0.011$ & (1) \\
12.1--18.1 & DCN        & 1.056  & $<0.018$ & (1) \\
11.9--18.1 & HC$_3$N    & 1.056  & $<0.025$ & (8) \\
11.9--18.1 & C$_2$H$_3$CN & 1.056  & $<0.1$ & (40) \\
11.9--18.1 & C$_2$H$_5$CN & 1.056  & $<0.1$ & (74) \\
11.9--18.1 & CH$_3$NH$_2$ & 1.056  & $<1.1$ & (27) \\
12.0--18.1 & HNCO       & 1.056  & $<0.07$ & (3) \\
11.9--18.1 & HCOOH      & 1.056  & $<0.28$ & (16) \\
11.9--18.1 & CH$_3$OCHO & 1.056  & $<1.1$ & (6) \\
11.9--18.1 & CH$_2$CO   & 1.056  & $<0.27$ & (9) \\
11.9--18.1 & CH$_2$OHCHO& 1.056  & $<0.33$ & (11) \\
11.9--18.1 & CH$_3$OCH$_3$& 1.056  & $<1.0$ & (7) \\
11.9--18.1 & c-C$_2$CH$_4$O & 1.056  & $<0.23$ & (6) \\
11.9--18.1 & CH$_3$COCH$_3$& 1.056  & $<0.32$ & (95) \\
11.9--18.1 & CH$_3$COOH & 1.056  & $<0.70$ & (58) \\
11.9--18.1 & Glycine I  & 1.056  & $<3.55$ & (33) \\
11.8--17.8 & PH$_3$     & 1.056  & $<1.2$ & (1) \\
11.8--17.8 & PN         & 1.056  & $<0.02$ & (1) \\
11.8--17.8 & PO         & 1.056  & $<0.11$ & (4) \\
17.1       & H$_2$O     & 1.057  & $<5100$ & (1) \\
11.8--18.1 & HDO        & 1.056  & $<1.4$ & (3) \\
\hline
\end{tabular}
\end{center}
\tablefoot{
  \tablefoottext{a}{Number of lines used for the determination of $Q$,
    in parentheses when individual lines are not clearly detected.}\\
  \tablefoottext{b}{Where we assume that HNC is produced in the
  coma with a Haser parent scale length of 1000~km \citep{Cor17}.}\\
  \tablefoottext{c}{Where we assume that H$_2$CO is produced in the
  coma with a Haser parent scale length of 5000~km.}
}
\end{table}

\subsection{Upper limits}
\label{sect-upperlimits}
Table~\ref{tabqpmoy} lists a large number of upper limits on the
production rate of various molecules of interest. Besides the
detection of some complex organic molecules \citep{Biv19},
these observations were
also the most sensitive ones ever undertaken in a Jupiter-family
comet. In most cases, we only took into account collisions and
radiative decay to compute the excitation. We did
not take into account infrared pumping via the rotational bands.
The comet was not very active, so the collision region was not
large, but due to its proximity to the Earth (0.08 au), the
IRAM beam was still probing regions (within 1000 km) unlikely
to be affected by infrared pumping. As discussed in
Sect.~\ref{sect-lifetime}, we probed molecules younger than
about 2500~s so that any radiative (IR-pumping, photo-destruction)
process that takes longer (g-factor, $\beta_0$ $<4\times10^{-4}$s$^{-1}$)
would not significantly affect the expected signal and derived production rate.
The corresponding upper limits in abundances relative to water
are given in Table~\ref{tababund}.

\subsection{Relative abundances}
\label{sect-abundances}
Table~\ref{tababund} provides the mean abundances relative to water or
upper limits derived from this study. Table~\ref{tababund} also lists
values derived in other comets from ground-based observations
\citep[e.g.][]{Biv19} and from in situ observations
in comet 67P. Abundances in comet 67P must be taken with caution when comparing
to those measured in comet 46P/Wirtanen since the technique was often very
different. In comet 67P, for some of the main species (e.g. CO, HCN, CH$_3$OH,
H$_2$S) \citet{Lau20} or \citet{Biv19b} derived relative bulk
abundances from the integration of molecular loss over two years.
They also use different techniques:
numerous local samplings by mass spectroscopy (with the Rosetta Orbiter
Spectrometer for Ion and Neutral Analysis -- ROSINA) 
vs. sub-millimetre mapping of the whole coma of some transitions (with the
Microwave Instrument for the Rosetta Orbiter -- MIRO), and
they show some discrepancies. For species of lower abundances,
mass spectroscopy with
ROSINA is mostly based on specific observing campaigns at specific times
\citep[e.g. before perihelion,][]{Rub19} and may not be representative
of the bulk abundance of the species. Most complex organic molecules
seem to be depleted in comet 67P, which could be due to this observational
bias, or it could be more specific to 67P.
Comet 46P has a composition quite comparable to most comets,
although some species such as HC$_3$N, HNCO, HNC, CS, and SO$_2$
seem to be relatively under-abundant.

\begin{table*}
\caption[]{Molecular abundances.}\label{tababund}
\begin{center}
\begin{tabular}{llllll}
\hline\hline
Molecule & Name &\multicolumn{3}{c}{Abundance relative to water in \%} \\  
         &      & in 46P & in comets & in 67P\tablefootmark{a} \\
\hline
HCN         & hydrogen cyanide & $0.11\pm0.01$   & 0.08--0.25   & 0.20 \\
HNC      & hydrogen isocyanide & $0.0015\pm0.0004$ & 0.002-0.035  & - \\
HNC\tablefootmark{b}      &    & $0.004\pm0.001$  &             &    \\
CH$_3$CN     & methyl cyanide  & $0.017\pm0.001$ & 0.008-0.054  & 0.0059 \\
HC$_3$N      & cyanoacetylene  & $<0.003$        & 0.002-0.068  & 0.0004 \\
HNCO         & isocyanic acid  & $<0.009$        & 0.009-0.080  & 0.027 \\
NH$_2$CHO    & formamide       & $0.015\pm0.002$ & 0.016-0.022  & 0.004 \\
\hline
CO           & carbon monoxide & $<1.23$         & 0.4- 35      &  0.3-3    \\ 
H$_2$CO$_{\rm ext}$\tablefootmark{c} & formaldehyde & $0.38\pm0.02$ & 0.13- 1.4    &  0.5 \\
CH$_3$OH     & methanol        & $3.38\pm0.03$   & 0.7 - 6.1    &  0.5-1.5   \\
HCOOH        & formic acid     & $<0.035$        & 0.03--0.18   &  0.013 \\
CH$_3$CHO    & acetaldehyde    & $0.06\pm0.01$   & 0.05--0.08   & 0.047\tablefootmark{d}\\
(CH$_2$OH)$_2$& ethylene glycol& $0.19\pm0.03$   & 0.07--0.35   & 0.011 \\
HCOOCH$_3$   & methyl formate  & $<0.14$         & 0.06--0.08   & 0.0034\tablefootmark{e}\\
CH$_2$OHCHO  & glycolaldehyde  & $<0.041$        & 0.016--0.039 & 0.0034\tablefootmark{e}\\
C$_2$H$_5$OH & ethanol         & $0.11\pm0.04$   & 0.11--0.19   & 0.10\tablefootmark{f}\\
\hline
H$_2$S      & hydrogen sulphide  & $0.92\pm0.01$  & 0.09- 1.5    & 2.0 \\
CS          & carbon monosulphide& $0.028\pm0.003$ & 0.05--0.20  & 0.02\tablefootmark{h}\\
OCS         & carbonyl sulphide  & $<0.07$        & 0.05--0.40   & 0.07 \\
SO          & sulphur monoxide   & $<0.09$        & 0.04--0.30   & 0.071 \\
SO$_2$      & sulphur dioxide    & $<0.03$        & 0.03--0.23   & 0.127  \\
H$_2$CS     & thioformaldehyde  & $\leq0.016$    & 0.009--0.090 & 0.0027 \\
NS          & nitrogen sulphide  & $<0.012$       & 0.006-0.012  &        \\
\hline
CH$_3$SH     & methyl mercaptan & $<0.06$     & $< 0.023$      & 0.038 \\ 
CH$_2$CO       & ketene         & $<0.034$    & $\leq 0.0078$  &  \\
CH$_3$COCH$_3$ & acetone        & $<0.04$     & $\leq 0.011$   & 0.0047\tablefootmark{g}\\
CH$_3$OCH$_3$  & dimethyl ether & $<0.13$     & $< 0.025$      & 0.04\tablefootmark{f} \\
c-C$_2$H$_4$O  & ethylene oxide & $<0.029$    & $< 0.006$      & 0.047\tablefootmark{d}\\
CH$_3$COOH     & acetic acid    & $<0.09$     & $< 0.026$      & 0.0034\tablefootmark{e}\\
CH$_3$NH$_2$   & methylamine   & $<0.13$     & $< 0.055$      &  \\
C$_2$H$_3$CN   & acrylonitrile  & $<0.013$    & $< 0.0027$     &  \\
C$_2$H$_5$CN   & ethyl cyanide  & $<0.013$    & $< 0.0036$     &  \\
PH$_3$         & phosphine      & $<0.15$     & $< 0.07 $      & $<0.003$ \\
PN             & phosphorus nitride & $<0.003$  & $< 0.0  $    & $<0.001$ \\
PO             & phosphorus oxide   & $<0.013$  & $< 0.0  $    & 0.011  \\
NH$_2$CH$_2$COOH & glycine I    & $<0.5$      & $< 0.18$       & 0.000017  \\
\hline
\end{tabular}
\end{center}
\tablefoot{
  \tablefoottext{a}{Based on \citet{Biv19,Rub19,Lau20}.} \\
  \tablefoottext{b}{Assuming a daughter distribution with $Lp=1000$ km \citep{Cor14,Cor17}.}\\
  \tablefoottext{c}{Assuming a daughter distribution.}\\
  \tablefoottext{d,e,f,g}{Isomers not distinguished by ROSINA; abundance is for their sum.}\\
   \tablefoottext{h}{From CS$_2$ with ROSINA.}
}
\end{table*}

\subsection{Isotopic ratios}
\label{sect-isotopes}
\begin{figure}[]
\centering
\resizebox{\hsize}{!}{\includegraphics[angle=270]{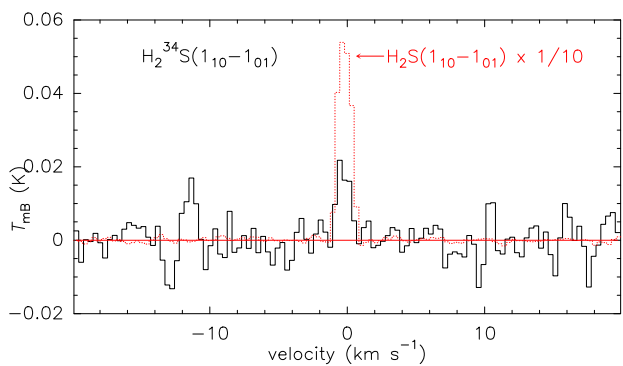}}
\caption{H$_2^{34}$S($1_{10}-1_{01}$) line at 167910.516~MHz together with
  the main isotopologue H$_2^{32}$S line divided by 10 and observed in the
  same FTS spectrum between 13.0 and 17.0 Dec. 2018.
  The vertical scale is the main beam brightness temperature,
  and the horizontal scale is the Doppler velocity in the comet rest frame.
  The feature at -11\kms~ does not correspond to any known molecular line
  and could be spurious.}
\label{figh234s}
\end{figure}

The proximity to the Earth of comet 46P/Wirtanen offered a rare opportunity
to attempt measurements of isotopic ratios in a Jupiter-family comet.
However, limited information was obtained due to the relatively low activity
level of the comet.
The derived upper limit for the deuterium-to-hydrogen ratio (D/H) in water
(Table~\ref{tabisotop})
is five times higher than the D/H value of $1.61\pm0.65\times10^{-4}$ measured
with SOFIA from the detection of the 509~GHz HDO line \citep{Lis19}.
Only H$_2^{34}$S could be clearly detected (Fig.~\ref{figh234s}).
The $^{32}$S/$^{34}$S in H$_2$S  and the limits
obtained in comet 46P for other isotopologues, with comparison to other
references, are provided in Table~\ref{tabisotop}.
The $^{32}$S/$^{34}$S in H$_2$S is compatible with the terrestrial value.
HC$^{15}$N shows a marginal $3-\sigma$ signal both in the FTS and VESPA
backends leading to $^{14}$N/$^{15}$N=$77\pm26$.
This is significantly lower than the mean value
\citep[$\approx150$,][]{Biv16}
observed in cometary HCN, suggesting that this signal could be spurious.

\begin{table*}
\caption[]{Isotopic ratios.}\label{tabisotop}
\begin{center}
\begin{tabular}{llllll}
\hline\hline
Ratio & Molecule & Value in 46P & other comets\tablefootmark{a} & in 67P\tablefootmark{b} & on Earth \\  
\hline
$^{12}$C/$^{13}$C & HCN     & $>66$          & 88--114  &        & 89.4 \\
$^{14}$N/$^{15}$N & HCN     & $>84$          & 139--205  &        & 272 \\
$^{32}$S/$^{34}$S & H$_2$S  & $20.6\pm2.9$   & 16--23    & 23.3   & 22.7 \\
$^{32}$S/$^{33}$S & H$_2$S  & $>21$          & --        & 151    & 126.9 \\
D/H              & HCN     & $<0.02$        & 0.0023    &        & \\
                 & H$_2$S  & $<0.029$       & $<0.008-<0.017$  & 0.0012 & \\
                 & H$_2$O  & $<8.8\times10^{-4}$\tablefootmark{c} &  &  & \\
                 & H$_2$O  & $1.6\pm0.6\times10^{-4}$\tablefootmark{d} & $1.4-6.5\times10^{-4}$ &  $5.3\times10^{-4}$ & $1.56\times10^{-4}$ \\
\hline
\end{tabular}
\end{center}
\tablefoot{
  \tablefoottext{a}{\citep{Boc15,Biv16,Cor19}}\\
  \tablefoottext{b}{\citep{Alt18}}\\
  \tablefoottext{c}{This paper, IRAM~30m upper limit.}\\
  \tablefoottext{d}{From SOFIA observations \citep{Lis19}.}
}
\end{table*}

\section{Discussion and conclusion}
\label{sect-discussion}
\subsection{Comparison with other observations}
Comet 46P/Wirtanen was the target of a worldwide observing campaign in
December 2018.\footnote{https://wirtanen.astro.umd.edu/}
It was observed with other millimetre-to-sub-millimetre
facilities such as JCMT \citep{Cou20}, ALMA \citep{Rot20,Biv21}, and
the Purple Mountain Observatory \citep{Wan20}, which
yield similar abundances. The short-term
variation of the methanol line shifts and intensity, likely related to
nucleus rotation, are observed in ALMA data \citep{Rot20,Rot21}.

Other observations \citep{Mou19,Han19} suggest an excited rotation
state of the nucleus with a primary period
around 9~h at the time of the observations.
The temporal sampling of our data set does not allow us to investigate
time variations related to nucleus rotation.
Infrared observations of comet 46P/Wirtanen
\citep{Del19,Bon20,Kha20} with the IRTF and Keck Telescope were obtained at
similar times: they find similar methanol production rates, and that,
compared to mean abundances measured among comets, 46P is relatively
methanol rich, while it presents typical abundances of
C$_2$H$_2$, C$_2$H$_6,$ and NH$_3$.

\subsection{Comparison with other comets}
Table~\ref{tababund} gives the abundances relative to water, or upper limits
we derived in comet 46P/Wirtanen and a comparison with values measured in all
other comets from millimetre-to-sub-millimetre ground-based observations
and values
measured in situ in comet 67P. Besides the differences in abundances relative
to water between 46P and 67P due to different techniques
(Sect.~\ref{sect-abundances}), the abundances found in comet 46P match most of
those measured in other comets. However, some species seem rather depleted:
CO (as is often observed in short period comets),
HNCO, HCOOH, HNC, CS, and SO$_2$.
This is not surprising for HNC, which is produced by a distributed source
and therefore rarefied within small fields of view \citep{Cor14}.
Assuming that it comes from a distributed source with a parent scale length of
1000~km \citep{Cor14,Cor17}, we find a HNC/HCN of 3.3\%, $2.7\times$ higher,
but still relatively low in comparison to other comets observed at $r_h$=1 au.
HNC was not detected in comet 73P-B \citep{Lis08}, which was observed at a
similarly small geocentric distance: the same interpretation could apply.
Although we assumed that CS is produced from the photo-dissociation of CS$_2$,
its abundance is still lower than in other comets, as suggested
in \citet{Cou20}.
The presence of another
distributed source of CS, which was proposed to explain the variation of
its abundance with heliocentric distance \citep{Biv11} could explain this
low abundance.
Likewise, a possible interpretation for the HCOOH depletion in the inner
coma of 46P is its production in the coma from the degradation of
the ammonium salt NH$_4^+$HCOO$^-,$ which was observed in 67P \citep{Alt20}.

\subsection{Conclusion}
We undertook a sensitive survey at millimetre wavelengths of
the Jupiter-family comet 46P/Wirtanen
in December 2018 at its most favourable apparition.
This allowed us to detect 11 molecules and obtain sensitive upper limits on
24 other molecules. This is the most sensitive millimetre survey of a
Jupiter-family comet regarding the number of measured molecular abundances.

Because the observations probed the inner coma ($<1000$~km), the inferred
abundances are weakly affected by uncertainties in molecular radiative
and photolysis lifetimes.
The abundances of complex molecules in 46P (e.g. CH$_3$CHO,
(CH$_2$OH)$_2$, NH$_2$CHO, C$_2$H$_5$OH) are similar to values
measured in long period comets. However, CH$_3$OH is more abundant.
C$_2$H$_5$OH/CH$_3$OH is lower (3\%) than in C/2014~Q2 (5.3\%),
C/2013~R1 ($\sim6.5$\%), or Hale-Bopp ($\sim8$\%).

Some molecules are particularly under-abundant in comet 46P, with
CO <1.2\%, which is often the case in JFCs, HNC, HNCO, and HCOOH.
A likely explanation for the depletion of these acids in the inner
coma of 46P is a production by distributed sources. There are
several lines of evidence, including maps, that HNC is produced by
the degradation of grains in cometary atmospheres \citep{Cor14,Cor17}.
Ammonium salts, including NH$_4^+$+HCOO$^-$, NH$_4^+$+OCN$^-,$
and NH$_4^+$+CN$^-$
have been identified in 67P's dust grains, impacting the
ROSINA sensor \citep{Alt20}.
Our measurements in 46P are consistent with HCOOH, HNCO, and HNC being
produced by the sublimation of ammonium salts.
The low CS abundance in 46P (0.03\% vs 0.05-0.20\% measured in other
comets 1 au from the Sun) suggests a distributed source of CS, other
than CS$_2$, in cometary atmospheres.
The investigation of isotopic ratios only yielded the measurements of the
$^{32}$S/$^{34}$S in H$_2$S, which is consistent with the terrestrial value.


\begin{acknowledgements}
This work is based on observations carried out under projects number
112-18 with the IRAM 30-m telescope and project number W18AB 
with the NOEMA Interferometer.
IRAM is supported by the Institut Nationnal des Sciences de l'Univers
(INSU) of the French Centre national de la recherche scientifique (CNRS),
the Max-Planck-Gesellschaft (MPG, Germany) and the Spanish IGN
(Instituto Geográfico Nacional).
We gratefully acknowledge the support from the IRAM staff for its
support during the observations.
The data were reduced and analysed thanks to the use of the GILDAS,
class software (http://www.iram.fr/IRAMFR/GILDAS).
This research has been supported by the Programme national de 
plan\'etologie de l'Institut des sciences de l'univers (INSU).
The Nan\c{c}ay Radio Observatory is operated by the Paris Observatory,
associated with the CNRS and with the University of Orl\'eans.
Part of this research was carried out at the Jet Propulsion Laboratory,
California Institute of Technology, under a contract with the National
Aeronautics and Space Administration. B.~P. Bonev and N. Dello Russo
acknowledge support of NSF grant AST-2009398 and NASA grant 80NSSC17K0705,
respectively.
N.X. Roth was supported by the NASA Postdoctoral Program at the NASA Goddard
Space Flight Center, administered by Universities Space Research Association
under contract with NASA. M.A. Cordiner was supported in part by the National
Science Foundation (under Grant No. AST-1614471). S.N. Milam and
M.A. Cordiner acknowledge the
Planetary Science Division Internal Scientist Funding
Program through the Fundamental Laboratory Research (FLaRe) work package,
as well as the NASA Astrobiology Institute through the Goddard Center for
Astrobiology (proposal 13-13NAI7-0032). M. DiSanti acknowledges support
through NASA Grant 18-SSO18\_2-0040.
\end{acknowledgements}


\clearpage
\onecolumn
\begin{appendix}
\section{Supplementary line list table}


\begin{longtable}{llllrrl}
\caption{\label{tabobs} Line intensities from IRAM and NOEMA observations.}\\ 
\hline\hline
Date & Molecule & Transition & Frequency & Pointing offset\tablefootmark{a} & Intensity\tablefootmark{b} & Doppler shift \\
$[yyyy/mm/dd.dd]$ &  &       & [GHz]     & [\arcsec] &       [K~\kms] & [\kms] \\
\hline
\endfirsthead
\caption{Continued.}\vspace{-0.3cm}\\
\hline\hline
Date & Molecule & Transition & Frequency & Pointing offset & Intensity & Doppler shift \\
$[yyyy/mm/dd.dd]$ &  &       & [GHz]     & [\arcsec] &       [K~\kms] & [\kms] \\
\hline
\endhead
\hline
\endfoot
\hline
\endlastfoot
2018/12/11.89 & HCN     & 3-2 &  265886.434 &  1.7 & $1.538\pm0.026$ & $-0.123\pm0.010$ \\
              &         &     &             &  6.1 & $1.221\pm0.034$ & $-0.196\pm0.017$ \\
              &         &     &             &  8.5 & $0.996\pm0.089$ & $-0.192\pm0.054$ \\
              &         &     &             & 12.0 & $0.863\pm0.064$ & $-0.234\pm0.047$ \\
              &         &     &             & 16.8 & $0.660\pm0.054$ & $-0.179\pm0.051$ \\
              &         &     &             & 20.6 & $0.366\pm0.082$ & $-0.264\pm0.124$ \\
              &         &     &             & 22.4 & $0.431\pm0.046$ & $-0.230\pm0.067$ \\
              &         &     &             & 26.4 & $0.375\pm0.086$ & $-0.160\pm0.164$ \\
\hline
\hline
\end{longtable}
\tablefoot{Frequencies and line parameters are from \citet{CDMS} and \citet{JPLmol}.\\
  \tablefoottext{a}{Mean pointing offset (for a series of lines).}\\
  \tablefoottext{b}{Whole line, including correction for hyperfine components not measured
    (e.g. HCN, H$^{13}$CN, DCN, H$_2^{33}$S).}
}
\newpage
\section{Full IRAM 30-m spectra of comet 46P/Wirtanen at 2~mm}
The following pages present the 147--185~GHz ($\lambda=2$mm)
FTS spectrum of comet 46P/Wirtanen
  obtained between 13.0 and 17.1 Dec. 2018 with the IRAM 30-m telescope.
  The three wavelength domains covered are plotted by series of $\approx2$~GHz
  windows with a spectral resolution smoothed to 0.78~MHz.
  Only spectra obtained within 2\arcsec~  of the nucleus were taken into
  account, and the strongest lines are identified.
  We note the much higher noise level around the frequency
  of the atmospheric H$_2$O line at 183310~MHz.
\begin{figure}[ht]\vspace{-1.0cm}
  \centering
\includegraphics[angle=0, width=16cm]{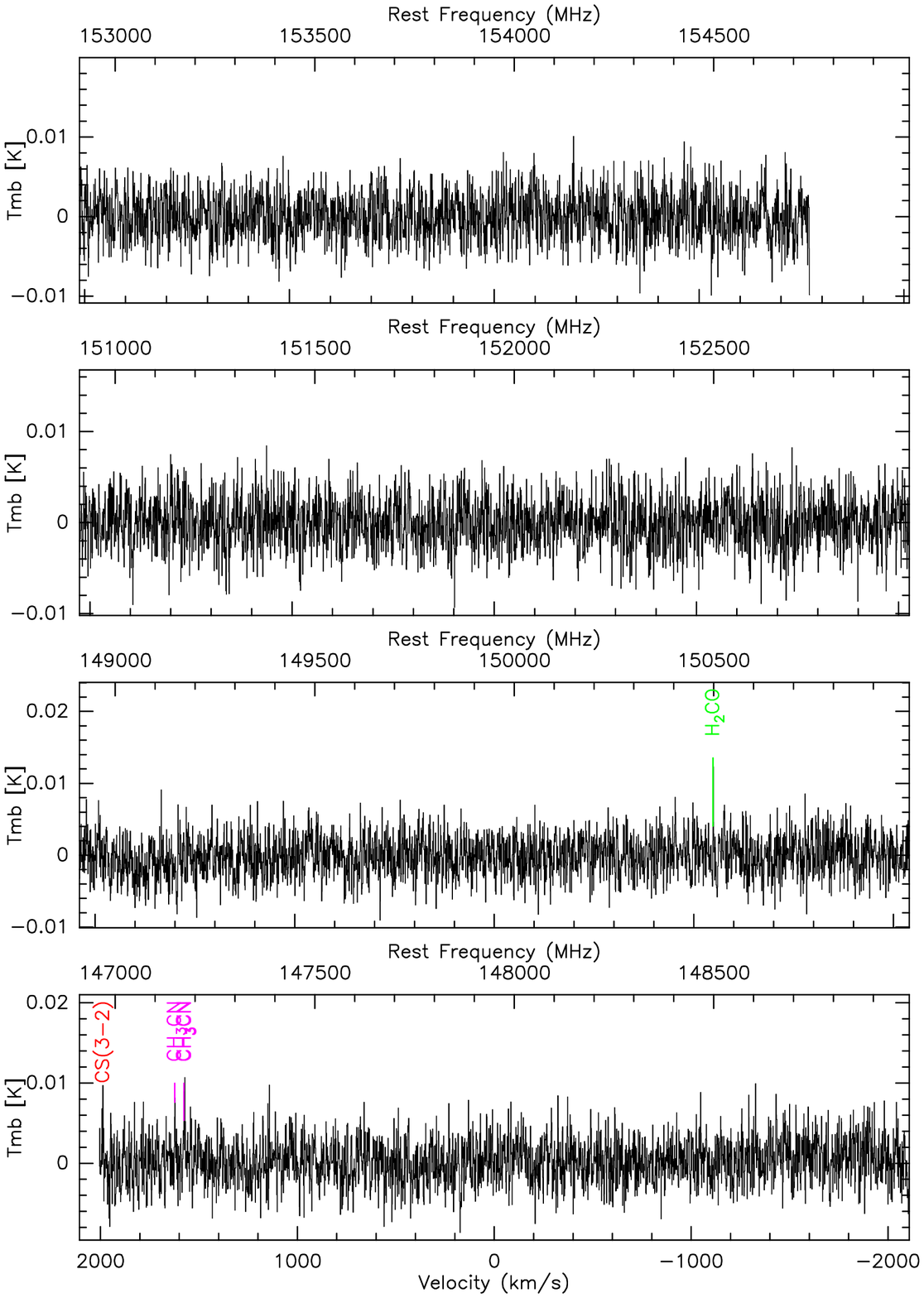}
\label{figsurvey2mm1}
\end{figure}
\begin{figure}[ht]\vspace{-0.0cm}
  \centering
\includegraphics[angle=0, width=16cm]{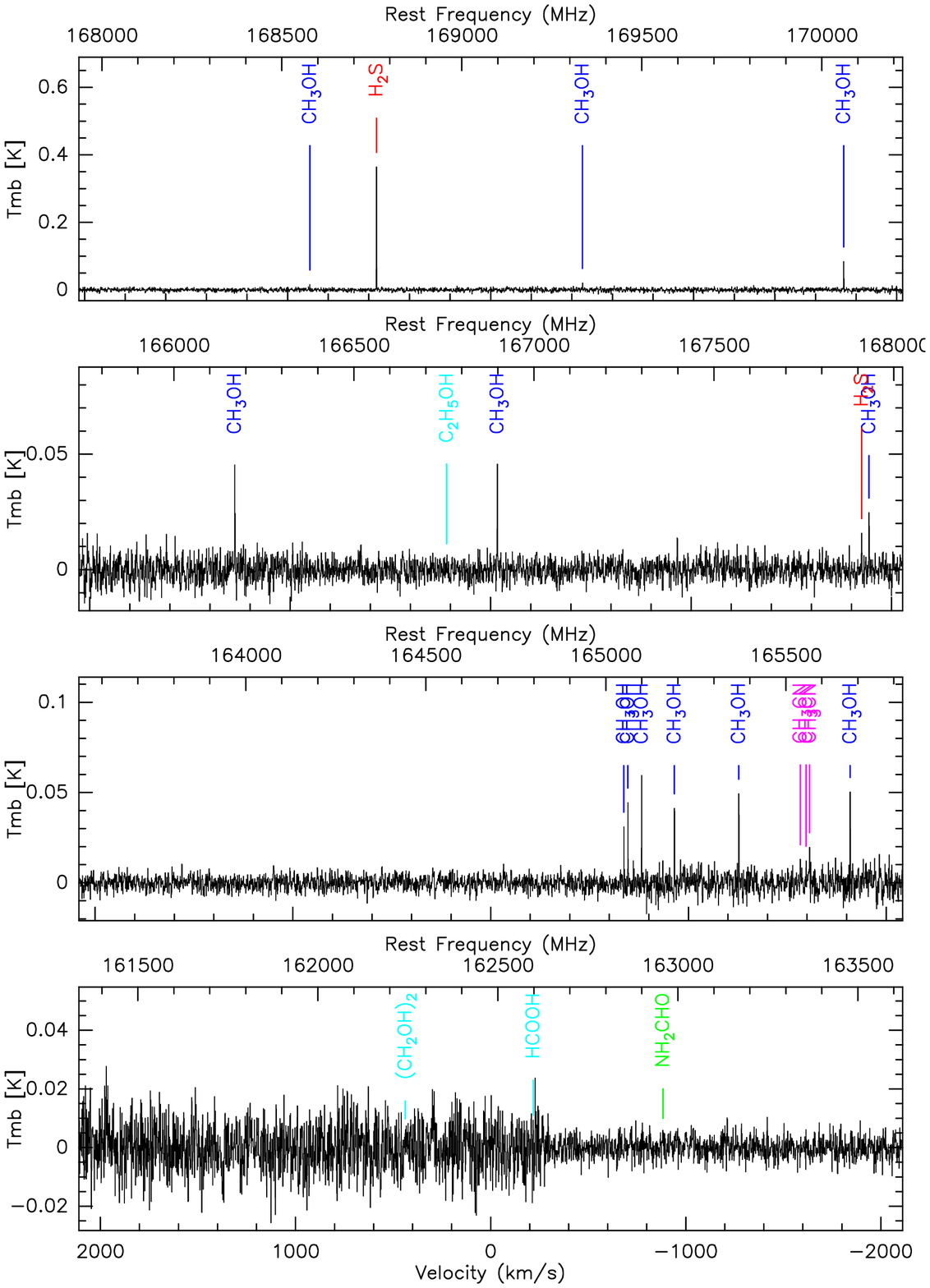}
\label{figsurvey2mm2}
\end{figure}
\begin{figure}[ht]\vspace{-0.0cm}
  \centering
\includegraphics[angle=0, width=16cm]{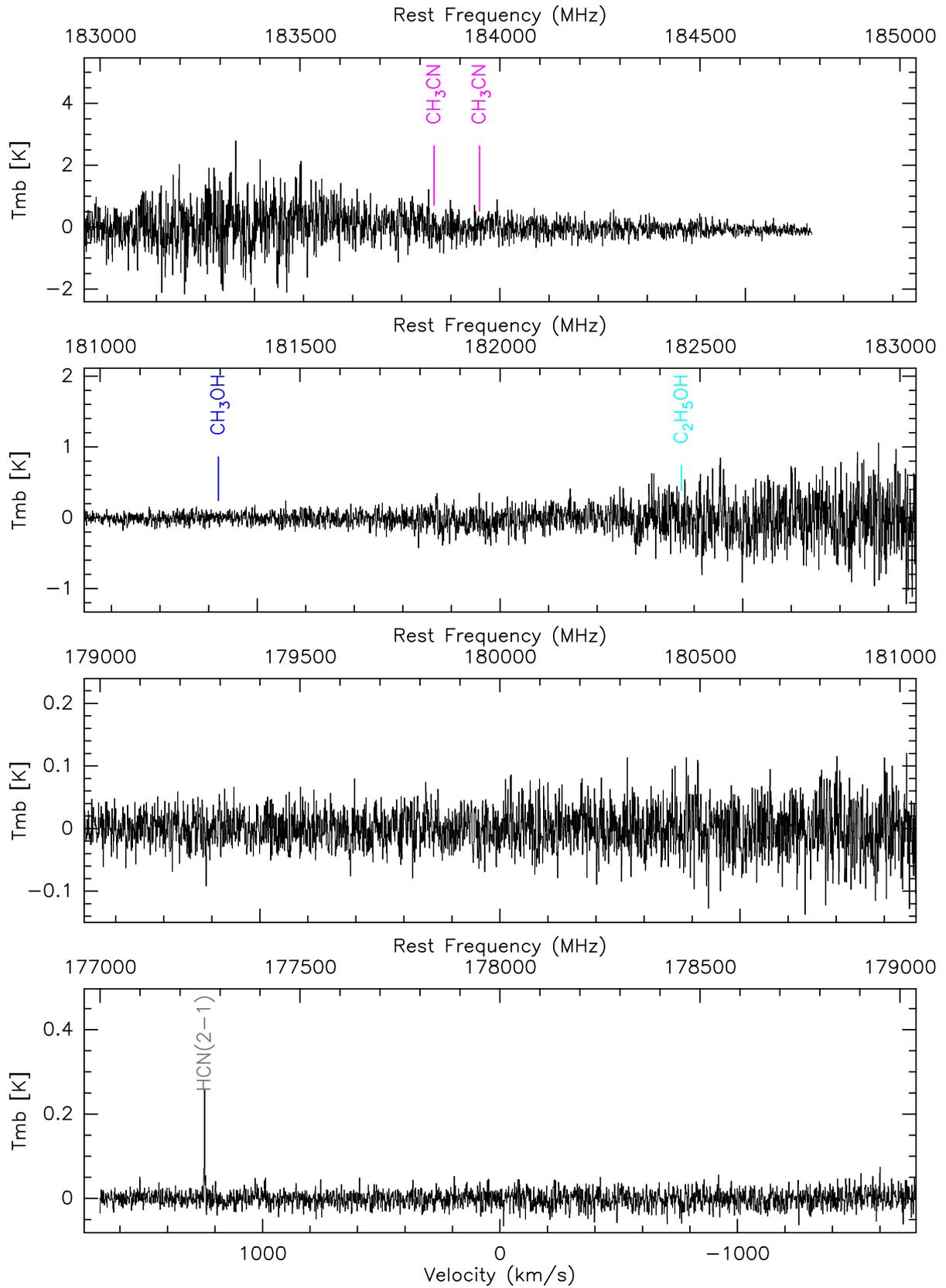}
\caption{
  Vertical scale in main beam brightness temperature adjusted to the lines
  or noise level.
  The frequency scale in the rest frame of the comet is indicated on the upper
  axis. A velocity scale with a reference at the centre of each band is indicated on
  the lower axis.
}
\label{figsurvey2mm3}
\end{figure}
\clearpage

\section{Full IRAM 30-m spectra of comet 46P/Wirtanen at 1~mm}
  The following pages present the 210--272~GHz ($\lambda=1.3$mm)
  FTS spectrum of comet 46P/Wirtanen
  obtained between 11.9 and 18.1 Dec. 2018 with the IRAM 30-m telescope.
  The 62~GHz, $\sim$320000 channel spectrum is plotted in $6\times4$ series of
  $\approx2.6$~GHz windows with a spectral resolution smoothed to 0.78~MHz.
  These spectra were obtained with four receiver tunings each covering
  $2\times7.78$~GHz. There are some overlaps and
  gaps in the final 62-GHz-wide frequency coverage.
  Only spectra obtained within 3\arcsec~ from the nucleus were taken into
  account.
  The strongest lines (signal $>2.5\times<\sigma>$) are labelled.
  Some identifications might be misleading where the
  local noise is higher than $<\sigma>$ and a noise peaks falls on a known
  molecular line resulting in a spurious detection.

\begin{figure}[ht]\vspace{-1.0cm}
  \centering
\includegraphics[angle=0, width=16cm]{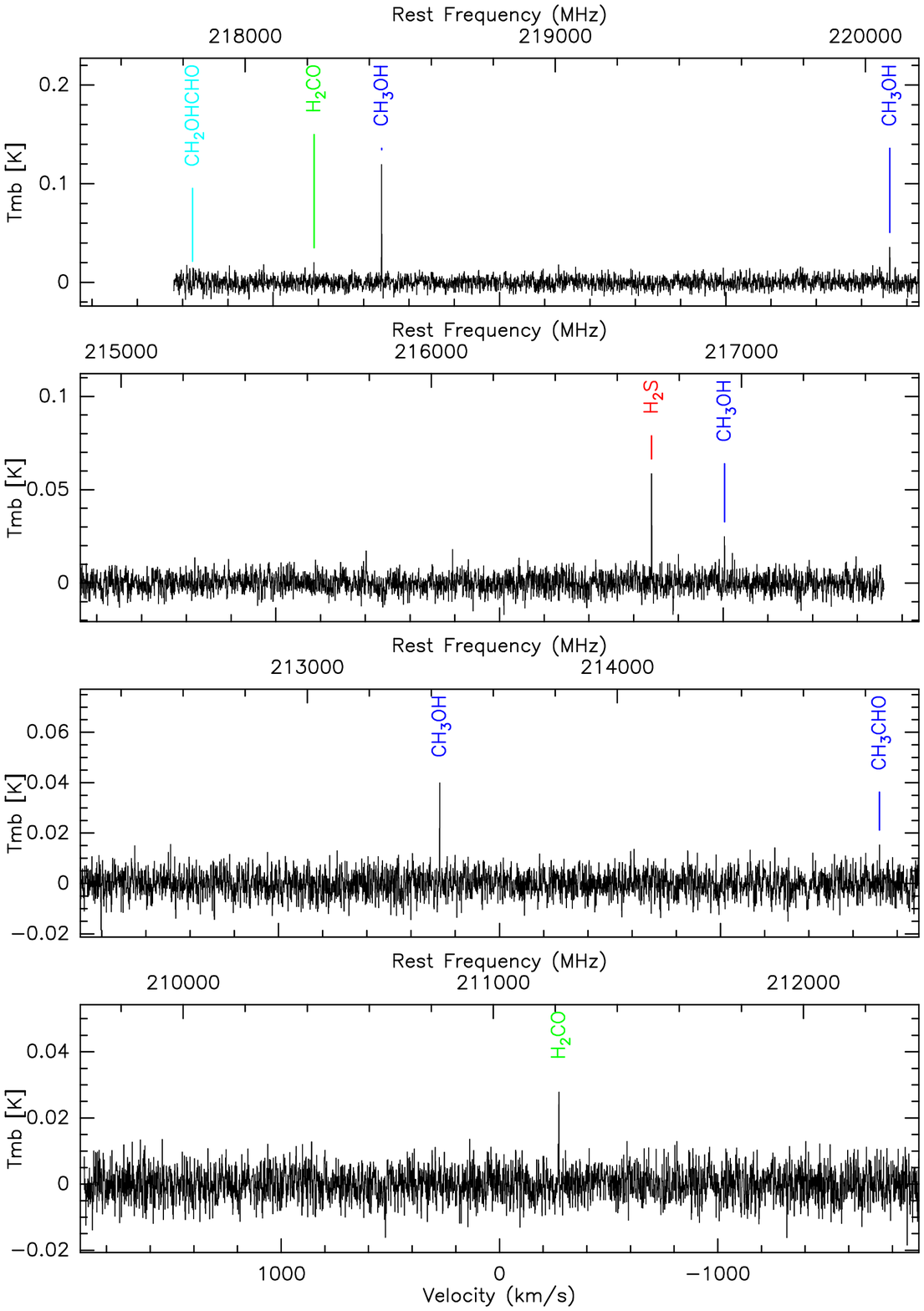}\vspace{-2.0cm}
\label{figsurvey1mm1}
\end{figure}
\begin{figure}[ht]\vspace{-0.0cm}
  \centering
\includegraphics[angle=0, width=16cm]{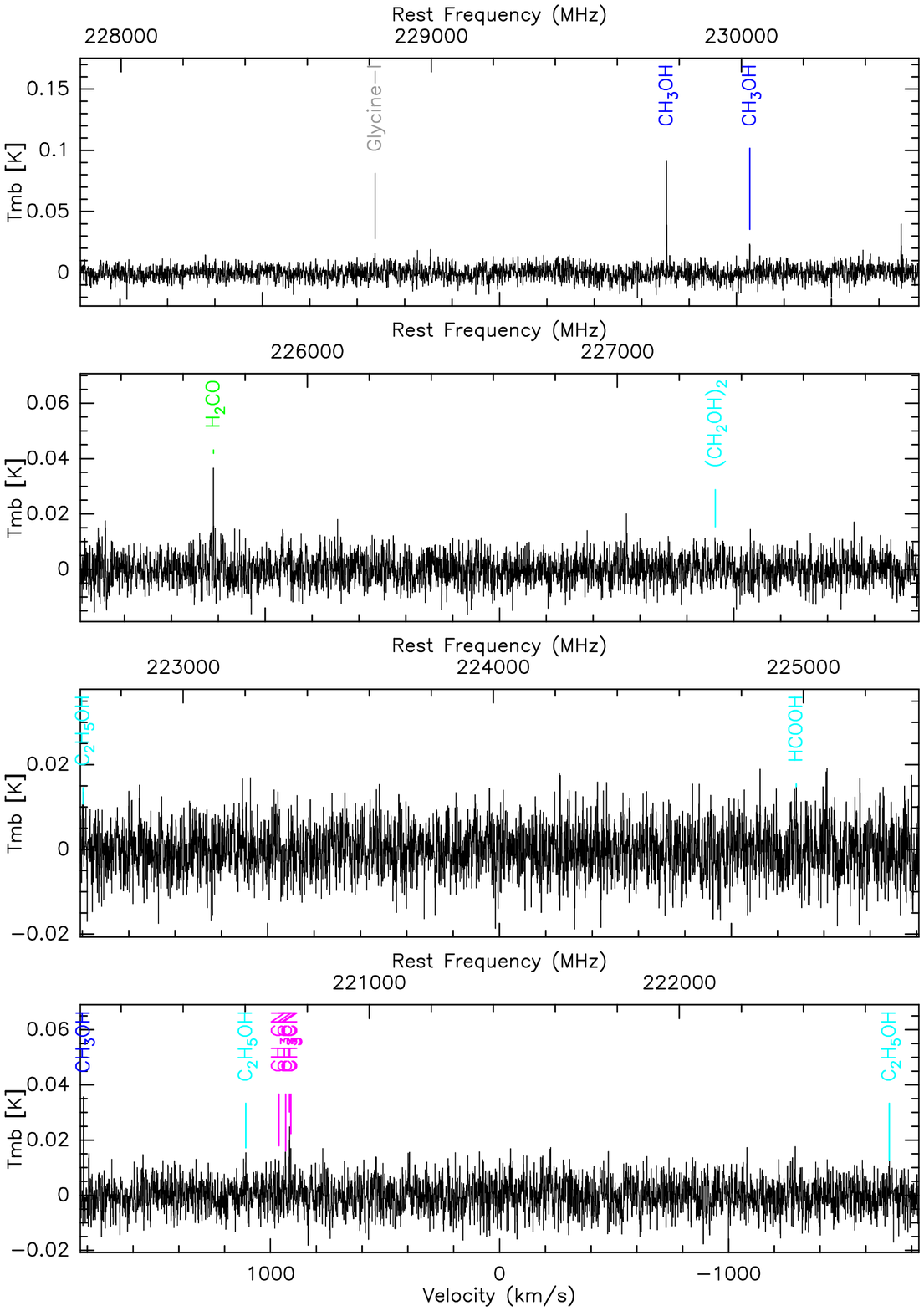}
\label{figsurvey1mm2}
\end{figure}
\begin{figure}[ht]\vspace{-0.0cm}
  \centering
\includegraphics[angle=0, width=16cm]{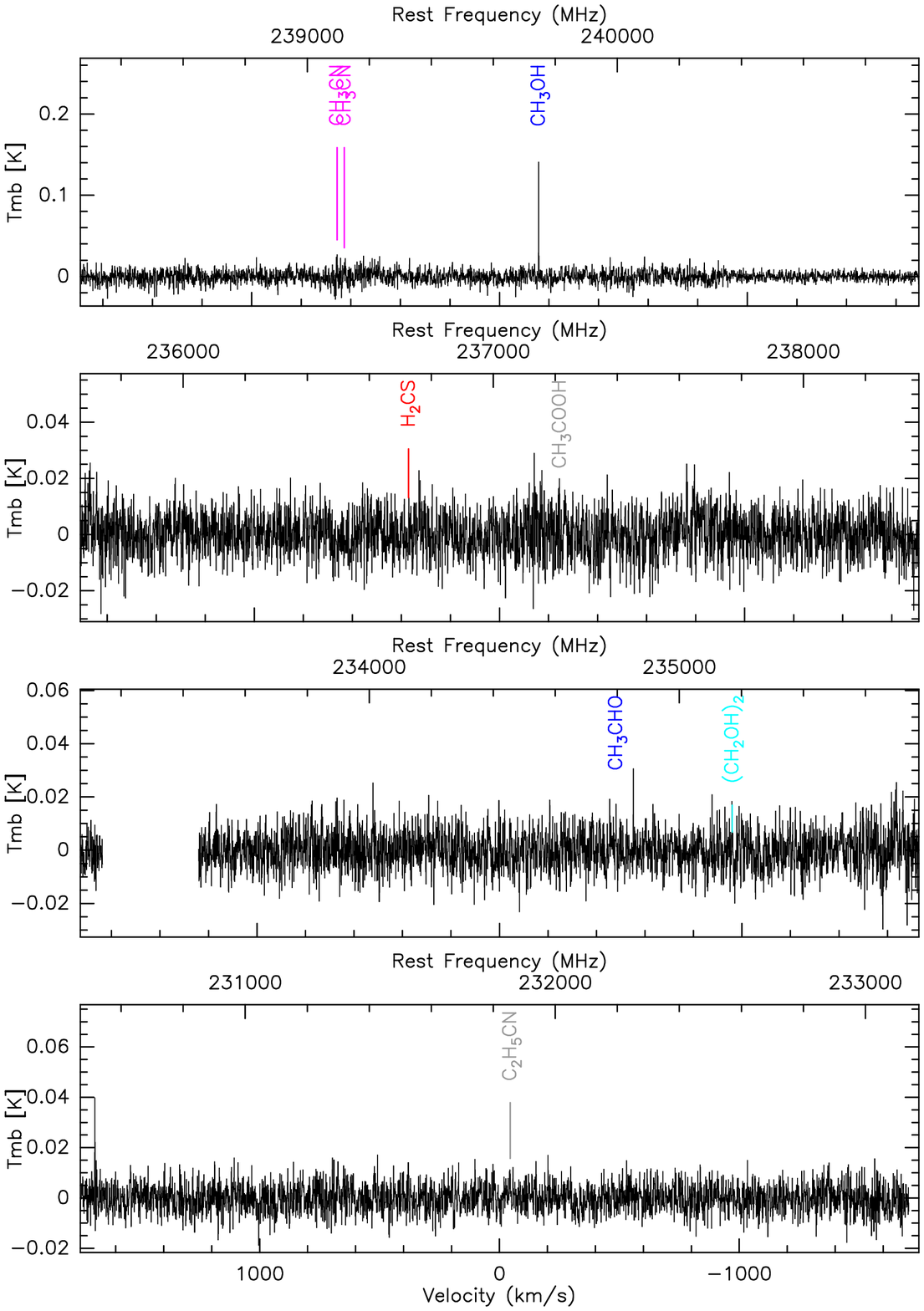}
\label{figsurvey1mm3}
\end{figure}
\begin{figure}[ht]\vspace{-0.0cm}
  \centering
\includegraphics[angle=0, width=16cm]{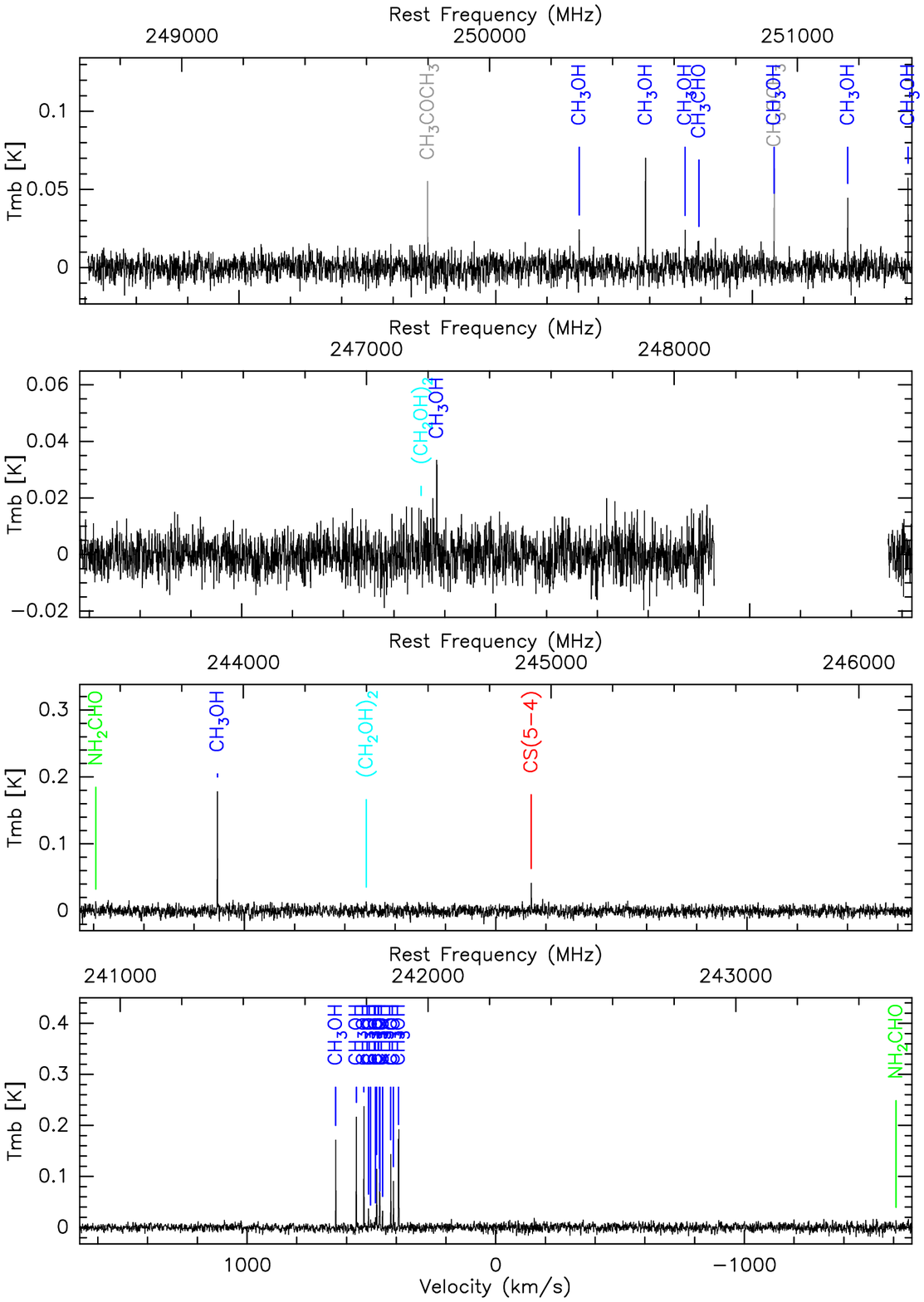}
\label{figsurvey1mm4}
\end{figure}
\begin{figure}[ht]\vspace{-0.0cm}
  \centering
\includegraphics[angle=0, width=16cm]{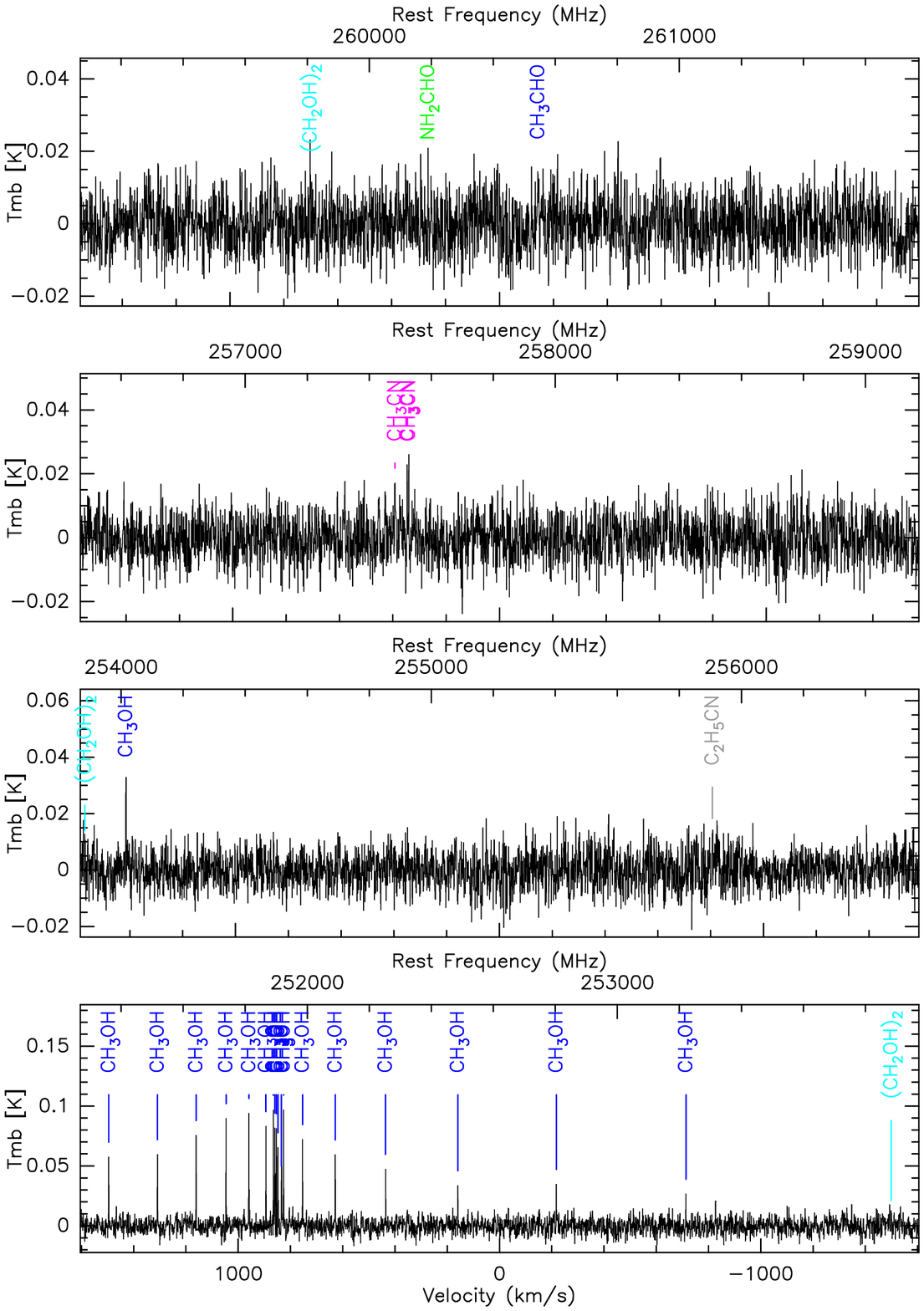}
\label{figsurvey1mm5}
\end{figure}
\begin{figure}[ht]\vspace{-0.0cm}
  \centering
\includegraphics[angle=0, width=16cm]{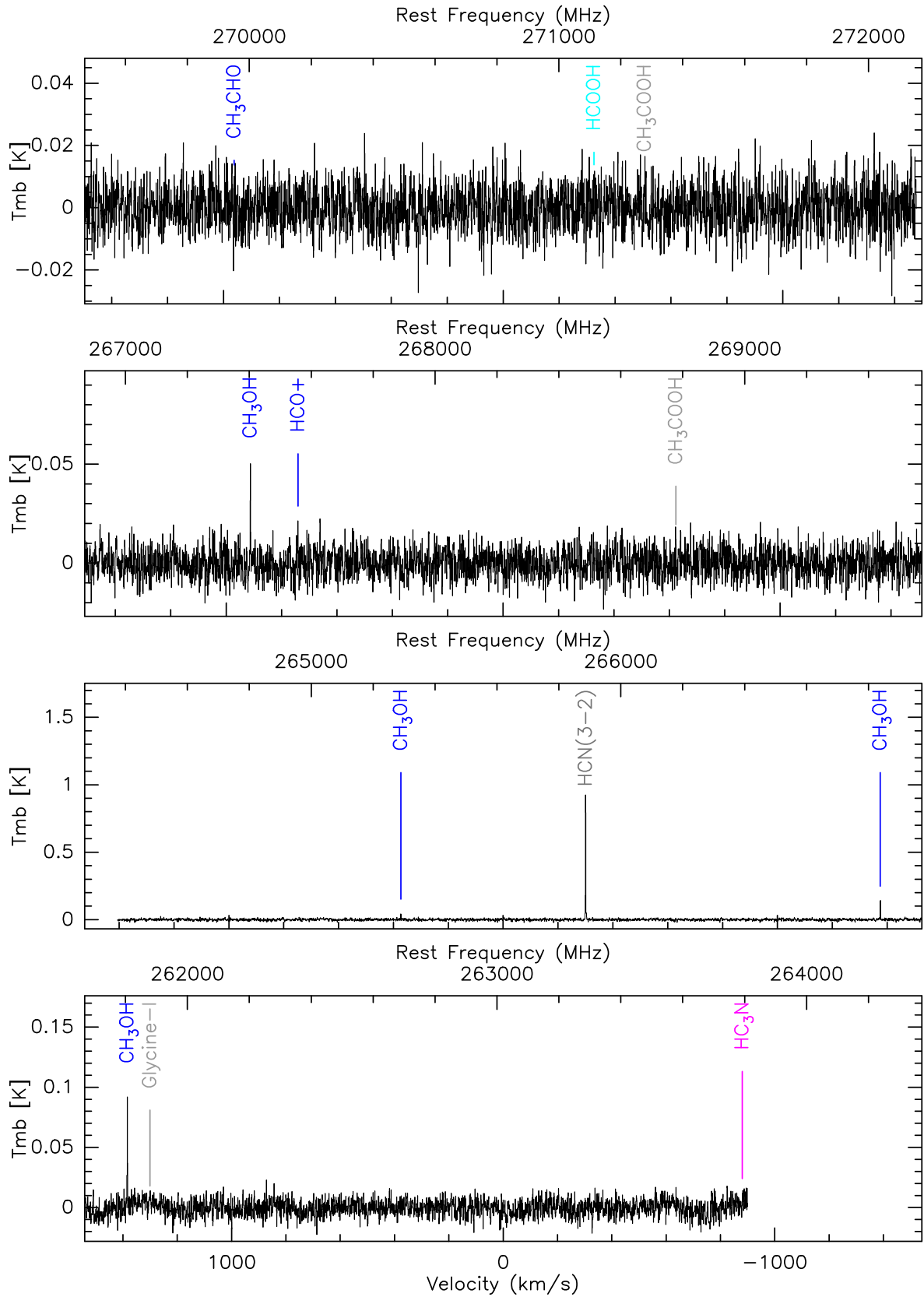}
\caption{Vertical scale in main beam
  brightness temperature adjusted to the lines or noise level.
  The frequency scale in the rest frame of the comet is indicated on the upper
  axis. A velocity scale with reference at centre of each band is indicated on
  the lower axis. The feature close to 230.5~GHz is due to contamination by
  galactic CO.
}
\label{figsurvey1mm6}
\end{figure}

\clearpage

\section{Selected molecular line widths in comets}
\label{app-linewidth}
Table~\ref{tablinewidth} provides the full width at half maximum ($FWHM$) or
velocity at half maximum ($VHM$) for asymmetric double peak lines, from Gaussian
fitting for molecular lines in comets. We took the average of several lines
when possible, and observations with similar beam sizes for a given comet,
in order to avoid being biased by spatial sampling.
When a (small) trend of line widths with beam size is observed, the width
has been interpolated to the mean beam size of the dataset.
The objective is to correlate the line width with the molecule lifetime
(when it is well known) or to constrain the lifetime of the molecule
from its line width (when it is poorly known), as described in
Section~\ref{sect-lifetime}.

\begin{table*}
\caption[]{Molecular line width in comets in m~s$^{-1}$}\label{tablinewidth}
\begin{center}
\begin{tabular}{llllll}
  \hline\hline
         & 46P:$FWHM$ & \multicolumn{3}{c}{Hale-Bopp $VHM$\tablefootmark{a}} & C/2014~Q2 $VHM$\tablefootmark{a} \\
\cline{3-5}
 Molecule\tablefootmark{b} & Dec. 2018 & Feb.1997\tablefootmark{c} & April 1997 & May 1997 & Jan. 2015 \\  
\hline
H$_2$S         & $1050\pm30$  &  $-920\pm70$ &  $-870\pm40$ & $-820\pm40$   & $-740\pm85$ \\
SO$_2$         & --           & $-1055\pm47$ &  $-962\pm40$ & $-1300\pm250$ & -- \\
H$_2$CS        & --           & --           & --           & --            & $-760\pm190$ \\
H$_2$CO        & $1260\pm100$ & $-1150\pm40$ & $-1030\pm57$ & $-965\pm8$   & $-855\pm17$ \\
SO             & --           & $-1103\pm37$ & --           & $-915\pm45$  & $-780\pm115$ \\
HCOOH          & --           &  $-980\pm50$ & --           & $-890\pm70$  & $-750\pm120$ \\
CH$_3$CHO      & $1230\pm240$ & $-1045\pm149$& --           & $-840\pm70$  & $-830\pm40$ \\
NH$_2$CHO      & $1200\pm198$ & --           &  $-890\pm56$ & $-542\pm165$ & $-900\pm70$ \\
OCS            & --           & $-1020\pm45$ & --           & $-890\pm90$  & $-840\pm90$ \\
HC$_3$N        & --           & $-1350\pm234$& $-1120\pm20$ & $-1060\pm60$ & $-670\pm140$ \\
(CH$_2$OH)$_2$ & $1105\pm186$ & --           & $-1030\pm200$& --           & $-770\pm60$ \\
HNCO           & --           & $-1260\pm62$ & --           & $-980\pm110$ & $-830\pm40$ \\
CS             & $1200\pm142$ & $-1120\pm40$ & $-1035\pm7$  & $-980\pm30$  & $-880\pm10$ \\
C$_2$H$_5$OH   & $1293\pm348$ & $-709\pm437$ & --           & --           & $-770\pm100$ \\
CH$_2$OHCHO    & --           & --           & --           & --           & $-1020\pm230$ \\
HCN            & $1165\pm10$  & $-1230\pm20$ & --           & $-1050\pm70$ & $-880\pm5$ \\
HNC            & $720\pm195$  & $-1230\pm15$ & --           & $-1200\pm110$ & $-1016\pm108$ \\
CH$_3$OH       & $1175\pm20$  & $-1320\pm30$ & $-1058\pm38$ &  $-960\pm30$  & $-867\pm4$ \\
CH$_3$CN       & $1030\pm118$ & --           & $-1014\pm144$& $-1000\pm80$  & $-760\pm50$ \\
CO             & --           & $-1270\pm60$ & $-1150\pm50$ &  $-960\pm90$  & $-810\pm70$ \\
\hline
\end{tabular}
\end{center}
\tablefoot{
 \tablefoottext{a}{Negative $VHM$ values correspond to the blueshifted
     side of the line (due to a sun-ward jet).}\\
 \tablefoottext{b}{Ordered by $\sim$increasing lifetime.}\\
 \tablefoottext{c}{Data obtained mostly with the 10.4~m Caltech Submillimeter
 Observatory or Plateau Bure 15~m antennas and a $24\pm4$\arcsec~ beamsize.
 April-May data were obtained with the IRAM 30~m and a $12\pm3$\arcsec~
 beamsize.}\\
 \tablefoottext{d}{CS, HNC, and more notably SO and H$_2$CO (which have
   a much shorter daughter/parent scale length ratio) are daughter
   molecules. Their scale length can be much longer than expected
   for a parent molecule.}
}
\end{table*}

\section{Dependence of the production rate on the uncertainty in photo-destruction rate}
\label{app-ncolbeta0}
\begin{figure}
\centering
\resizebox{\hsize}{!}{\includegraphics[angle=270]{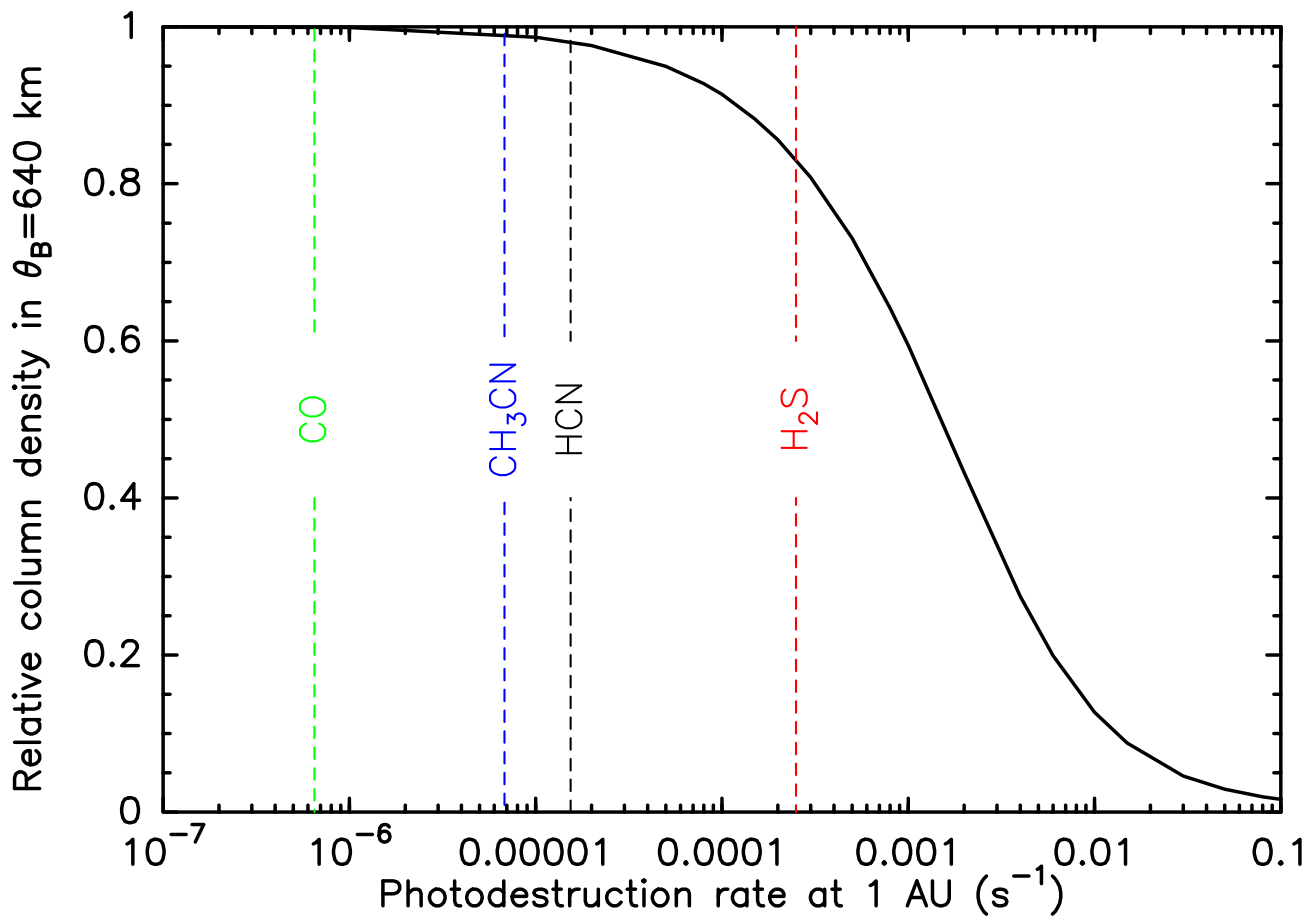}}
\caption{Plot showing evolution of the number of molecules in the
  IRAM 30-m beam as a function of molecule photo-destruction rate.
  We plot, more precisely, the average column density (number
  of molecules divided by the beam area) normalised to its value
  for infinite lifetime. We used the average beam size
  ($\theta_b$=11\arcsec) for an average pointing offset of 2\arcsec
  at a geocentric distance of 0.08 au, which is typical for most observations
  of comet 46P. A change of molecular photo-destruction rate will
  result in a change of the column density that can be inferred from this
  plot. The position of the photo-destruction rate for well-known molecules
  is indicated.
}
\label{figcolbeta0}
\end{figure}

 Figure~\ref{figcolbeta0} is provided here for the mean observing
 circumstances of the observations of comet 46P with IRAM 30-m telescope.
 In a first approximation, the line intensity is proportional to the
 column density $N_c$, which is itself proportional to the
 molecular production rate $Q$. So, for a given observed line intensity,
 a change of $N_c$ due to a change of the photo-destruction rate $\beta_0$,
 has to be compensated by an inverse change of
 $Q$: $\frac{\Delta Q}{Q} = -\frac{\Delta N_c}{N_c}$.
 
 From this plot, we can directly determine the change of the derived
 production rate $Q$ when we change the value $\beta_0$. For example, if
 $\beta_0$ is increased from 2 to $20\times10^{-5}$s$^{-1}$ , we get 
 $\frac{\Delta Q}{Q} = +0.12$, meaning the retrieved production rate
 will be increased by only 12\%. More generally,
 if the photo-destruction rate is not known but can be assumed to be
 less than that of H$_2$S, an error of less than 16\%  on the retrieved
 production rate is anticipated from these observations of comet 46P.

\end{appendix}

\end{document}